\documentclass[aps,preprint,preprintnumbers,nofootinbib,tightenlines]{revtex4}
\usepackage{graphicx}
\usepackage{amssymb}
\usepackage{amsmath,mathrsfs,verbatim}
\usepackage{times}
\usepackage{latexsym}
\def\beq{\begin{equation}}
\def\eeq{\end{equation}}
\def\bey{\begin{eqnarray}}
\def\eey{\end{eqnarray}}

\def\lsim{\mathrel{\raise.3ex\hbox{$<$\kern-.75em\lower1ex\hbox{$\sim$}}}}
\def\gsim{\mathrel{\raise.3ex\hbox{$>$\kern-.75em\lower1ex\hbox{$\sim$}}}}

\def\kms{\, {\rm km/s} }
\def\cmg{\, {\rm cm^2/g} }

\newcommand{\be}{\begin{equation}}
\newcommand{\ee}{\end{equation}}

\newcommand{\mx}{m_X}
\newcommand{\ax}{\alpha_X}
\newcommand{\mphi}{m_\phi}

\newcommand{\san}{\sigma_{\rm an}}

\begin{document}

%\preprint{MCTP/12-xx}

\title{Beyond Collisionless Dark Matter: \\ Particle Physics Dynamics for Dark Matter Halo Structure}

\author{Sean Tulin, Hai-Bo Yu, and Kathryn M. Zurek}
\affiliation{Michigan Center for Theoretical Physics, University of Michigan, Ann Arbor, MI 48109}

\date{\today}

\begin{abstract}

Dark matter (DM) self-interactions have important implications for the formation and evolution of structure, from dwarf galaxies to clusters of galaxies.  We study the dynamics of self-interacting DM via a light mediator, focusing on the quantum resonant regime where the scattering cross section has a non-trivial velocity dependence. While there are long-standing indications that observations of small scale structure in the Universe are not in accord with the predictions of collisionless DM, theoretical study and simulations of DM self-interactions have focused on parameter regimes with simple analytic solutions for the scattering cross section, with constant or classical velocity (and no angular) dependence.  We devise a method that allows us to explore the velocity and angular dependence of self-scattering more broadly, in the strongly-coupled resonant and classical regimes where many partial modes are necessary for the achieving the result.  We map out the entire parameter space of DM self-interactions --- and implications for structure observations --- as a function of the coupling and the DM and mediator masses.  We derive a new analytic formula for describing resonant $s$-wave scattering.  Finally, we show that DM self-interactions can be correlated with observations of Sommerfeld enhancements in DM annihilation through indirect detection experiments.  

\end{abstract}

\maketitle

\section{Introduction}

Dark Matter (DM) is five times as prevalent as ordinary matter, and yet its particle physics nature remains elusive. Efforts are underway to detect it through its non-gravitational interaction with ordinary matter via direct scattering off nuclei in underground experiments, annihilation to Standard Model (SM) by-products in the galaxy today, and direct production in terrestrial collider experiments.  There are well-motivated theoretical reasons to think that DM may reveal itself through these means, shedding light on the underlying theory of DM.  On the other hand, all evidence for DM thus far has been obtained through its gravitational interactions, and it remains important to continue exploring the nature of DM through its effects on structure in the Universe.

The formation of structure in the Universe gives critical information about the nature of the DM sector. As is well known, the collisionless cold DM (CCDM) paradigm has been highly successful in accounting for large scale structure of the Universe. However, it is  far from clear that this paradigm can also successfully explain the small scale structure of the Universe. Precision observations of dwarf galaxies show DM distributions with cores~\cite{Oh:2010ea}, in contrast to cusps predicted by CCDM simulations. It has also been shown that the most massive subhalos in CCDM simulations of Miky Way (MW) size halos are too dense to host the observed brightest satellites of the MW~\cite{BoylanKolchin:2011de,BoylanKolchin:2011dk}. Lastly, chemo-dynamic measurements in at least two MW dwarf galaxies show that the slopes of the DM density profiles are shallower than predicted by CCDM simulations~\cite{Walker:2011zu}. 

These small scale anomalies, taken at face value, may indicate that other interactions besides gravity play a role in structure formation. An interesting possibility is that DM carries self-interactions~\cite{Spergel:1999mh}. In this scenario, heat can be conducted from the hotter outer to the cooler inner parts of the halo through DM collisions, which softens the density profile in the central regions of the halo. Recent simulations show that the typical cross section needed to flatten the cores of galaxies is $\sigma \sim 10^{-24} \, {\rm cm}^2 \times (m_X/{\rm GeV})$~\cite{Vogelsberger:2012ku,Rocha:2012jg,Zavala:2012us}, where $m_X$ is the DM mass.  Since it is far larger than the typical weak-scale cross section, $\sigma \sim 10^{-36} \, {\rm cm}^2$, DM candidates cannot be usual WIMPs.   On the other hand, a light dark force, denoted $\phi$, can provide the required large cross section.  A perturbative calculation for the scattering cross section gives (in the small velocity limit)
\begin{equation}
\sigma \approx 5 \times 10^{-23} \, {\rm cm}^2 \left(\frac{\alpha_X}{0.01}\right)^2 \left(\frac{m_X}{10 \: {\rm GeV}}\right)^2 \left(\frac{10 \: {\rm MeV}}{m_\phi}\right)^4 \, ,
\label{cross-section}
\end{equation} 
where the coupling $\alpha_X$ is the DM analog of the fine structure constant. Eq.~\eqref{cross-section} shows that light dark force with electromagnetic strength coupling can ameliorate discrepancies in small scale structure observations.  Interestingly, light mediators exist in many DM models which are motivated to solve completely different problems~\cite{Pospelov:2007mp,Hooper:2008im,Feng:2008ya,Feng:2008mu, ArkaniHamed:2008qn, Pospelov:2008jd,Shepherd:2009sa,Kaplan:2009ag,An:2009vq}.

Light forces, mediated by a Yukawa potential, can have rich dynamics.  The DM self-scattering cross section may be velocity dependent, in contrast to the original model where a constant cross section is assumed~\cite{Spergel:1999mh}. In the regime where $\alpha_X m_X/m_\phi \gtrsim 1$, Eq.~\eqref{cross-section} breaks down  and the non-perturbative effect plays a key role in DM scattering. When the momentum transfer is much larger than the mediator mass, scattering occurs in the Coulomb limit and the cross section is proportional to $\sim1/v^4$ with $v$ as the DM relative velocity~\cite{Feng:2009mn,Ackerman:2008gi}. While in the quantum resonant regime, the scattering cross section can be enhanced and scale as~$1/v^2$ due to the formation of quasi-bound states~\cite{Tulin:2012wi}. This is the same mechanism that leads to resonant Sommerfeld enhancements in DM annihilation~\cite{Tulin:2012wi}. These features have important consequences for DM halo dynamics because scattering is enhanced on dwarf galaxy scales compared to MW and cluster scales. It provides a natural mechanism to evade constraints from large scales such as elliptical DM halo shapes and the Bullet Cluster.  It has been shown that most of parameter space of interest for thermal DM is in this non-perturbative regime~\cite{Tulin:2012wi}.

While self-interacting DM has been the subject of astrophysical interest and numerical simulation, the particle physics aspects have been comparatively little examined.  Studies have so far limited themselves to regions that can be approximated analytically through classical or Born formulae~\cite{Feng:2009hw,Ibe:2009mk,Loeb:2010gj,Lin:2011gj,Aarssen:2012fx}, or else have considered a limited range of parameter space~\cite{Buckley:2009in}.  The purpose of this paper is to study the full range of effects of light dark force dynamics on halo structure, and is intended as a companion paper to \cite{Tulin:2012wi}.  In \cite{Tulin:2012wi}, we have laid out a model of self-interacting DM that satisfies relic density considerations while giving rise to a rich structure in the scattering, including the presence of velocity dependent resonances.  Here we delve into many more details.  We discuss the method that we use for an improvement in the numerical efficiency for solving the Schr\"{o}dinger equation such that we are able to reach regions of scattering parameter space where many partial wave $\ell$ modes are required.  This method allows us to explore the strongly-coupled resonant and classical regimes.  We are able to verify numerically the classical formula, which has never been done before.  We are also able to examine the angular dependence of the scattering cross section in the classical and strongly-coupled regimes, observing the transition to the weakly coupled regime with forward-peaked Rutherford scattering.  We examine in detail $s$- and $p$-wave resonances in the strongly-coupled regime, and provide benchmark points for simulations.  

The outline of this paper is as follows.  First, we discuss the case for self-interacting DM, summarizing the current status of DM simulations and observations of small scale structure.  Then we describe our setup, diving into technical details of solving the dynamics of strongly-coupled systems.  We show the results of our method, encapsulating the Born, resonant, and classical regimes, and we examine the velocity and angular dependent scattering effects on halo structure.  We then present a new analytic result for the $s$-wave resonant regime before connecting our method to relic density calculations, explicitly including the effect of the Sommerfeld enhancement.  Lastly, we discuss connections to observation in indirect detection experiments, showing how the enhancement in self-scattering can also be important for DM annihilation.  We then conclude.  

%%%%%%%%%%%%%%%%%%%%%%%%%%%%%%%55
\section{Self-interacting dark matter and small scale structure}
%%%%%%%%%%%%%%%%%%%%%%%%%%%%%%%%%

For some time there has been debate about whether the paradigm of CCDM accurately describes the observed small scale structure in the Universe.  Small scale objects (e.g., dwarf galaxies) are typically DM-dominated, and therefore offer potentially cleaner laboratories to test CCDM predictions compared to systems with higher baryon densities.  Here, we describe three discrepancies, and show how two of them may point beyond CCDM in the form of DM self-interactions.  We emphasize, however, that the situation remains far from clear, and ultimately more detailed numerical simulations including baryonic effects are required before drawing definitive conclusions~\cite{Scannapieco:2011yd,Kuhlen:2012ft}.

{\it Core-vs-cusp problem:} The central density profiles of dwarf galaxy halos indicate a long-standing discrepancy between steep cusps predicted by CCDM-only simulations \cite{Dubinski:1991bm,Navarro:1996gj,Wechsler:2001cs} compared to flat cores inferred from observed galaxy rotation curves  \cite{Flores:1994gz,de Blok:2001fe,Simon:2004sr,Oh:2010ea,deNaray:2011hy}.   Observations of clusters of galaxies may also exhibit cored profiles~\cite{Sand:2003ng,Newman:2009qm,Newman:2012nw}.  Baryonic effects may provide an astrophysical mechanism for flattening the DM density profile in the center of a galaxy (or cluster of galaxies), which are often baryon dominated.  It has been argued that feedback from (dissipational) baryonic matter leads to further contraction of the central DM cusp~\cite{Blumenthal:1985qy}, further exacerbating the discrepancy.  However, simulations have shown this mechanism to be less effective than previously thought~\cite{Gnedin:2004cx,Tissera:2009cm}.  Moreover, supernova feedback may have the opposite effect: supernova energy injected into the interstellar medium leads to baryonic outflow, which can gravitationally disrupt the central cusp, resulting in lower DM densities compared to CCDM-only simulations \cite{Navarro:1996bv,Governato:2009bg,Oh:2010mc,Brook:2011nz,Pontzen:2011ty,Governato:2012fa}.  However, this mechanism seems unlikely to explain central cores in metal-poor galaxies with limited star formation rates~\cite{deNaray:2011hy}.

{\it Missing satellites problem:}  There has been an order of magnitude discrepancy between the number of observed and expected satellites of the Milky Way (MW) \cite{Kauffmann:1993gv,Klypin:1999uc,Moore:1999nt,Bullock:2010uy}. Baryonic processes such as supernova feedback and/or photoionization may play important for suppressing star formation in dwarf galaxies, explaining the observed (weak) baryon content of these small galaxies~\cite{Bullock:2000wn}.  Recently, the Sloan Digital Sky Survey has discovered many faint galaxies, such that it is evident that as many as a factor of $5-20$ of the known dwarf galaxies could be still undiscovered due to faintness, luminosity bias and limited sky coverage \cite{Tollerud:2008ze,Walsh:2008qn,Bullock:2009gv}.  Consensus is thus shifting toward the view that the number of MW subhalos is not an issue for the predictions of CCDM, at least for smaller subhalos.

{\it Too-big-to-fail problem:} 
Detailed studies of the brightest MW dwarf spheroidal (dSph) galaxies, which are DM dominated at all radii, show discrepancies with CCDM-only predictions (see e.g.~\cite{Walker:2012td}).  These satellites are expected to be hosted by the largest subhalos in the MW halo, as they have the largest velocity dispersions observed from their rotation curves.  However, the most massive subhalos predicted by CCDM-only simulations are too massive, with central densities too large, to host the brightest observed satellites~\cite{Strigari:2007ma,BoylanKolchin:2011de,BoylanKolchin:2011dk}.  Simulations predict $\mathcal{O}(10)$ subhalos with maximum circular velocity $V_{\rm max} > 30 \mbox{ km/s}$, whereas the MW dSphs have $V_{\rm max} < 25 \mbox{ km/s}$.\footnote{These predicted most massive subhalos are ``too big to fail'' in forming stars, unlike shallower potentials of much smaller subhalos.}  This discrepancy may share a common resolution with the core-vs-cusp problem; these massive subhalos can be reconciled with the observed dSphs if their central densities are reduced compared to CCDM predictions.  Indeed, analyses of stellar subpopulations within several dSphs indicate cored central density profiles~\cite{Kleyna:2003zt,SanchezSalcedo:2006fa,Goerdt:2006rw,Walker:2011zu,Jardel:2011yh} (except for Draco~\cite{Jardel:2012am}).  Several studies using hydrodynamical simulations have suggested that baryonic physics --- i.e., feedback from star formation and supernovae, as well as ram pressure and tidal stripping from the host halo --- may induce dSph cores~\cite{Mayer:2000ti,Sawala:2012cn,Zolotov:2012xd,Arraki:2012bu}, while Ref.~\cite{Parry:2011iz} found a smaller impact from baryonic effects.  Additionally, the severity of this problem can be reduced by taking into account statistical variation in the formation of MW-sized halos~\cite{Purcell:2012kd}, as well as uncertainty in the MW halo mass which sets the normalization of the subhalo mass spectrum.  Larger MW halo masses lead to a larger discrepancy between simulation and observed MW satellites.  For example, Ref.~\cite{Zolotov:2012xd} used a MW mass of $8 \times 10^{11} M_\odot$ and saw no too-big-to-fail problem.  On the other hand, Ref.~\cite{BoylanKolchin:2010ck} argued that a larger MW mass, around $2 \times 10^{12} M_\odot$, is warranted.  Whether this larger estimate or the smaller one advocated in \cite{Xue:2008se} prevails will have important implications for the too-big-to-fail problem.

Given these persistent questions about the accordance of observations with the predictions of CCDM, it is interesting to look beyond the paradigm of cold and collisionless DM.  One of the first attempts to do this was to give DM some kinetic energy, i.e., to make it warm.   Warm DM predicts a suppression in the halo mass function at small scales, below the free-streaming length.  Thus, warm DM effectively removes substructure, and predicts a reduced number of satellites in a galaxy such as the MW.  On the other hand, warm DM halos may be less concentrated than CCDM halos on scales of order the free-streaming length, but they are still cuspy \cite{VillaescusaNavarro:2010qy,Maccio:2012qf}.  As a result, warm DM solves only the missing satellites problem, which is considered the least severe discrepancy, but not the remaining problems.

The other known mechanism for changing the structure of DM halos is self-interactions.  Self-interacting DM was introduced as a solution to the core-vs-cusp and missing satellites problems in Ref.~\cite{Spergel:1999mh}.  Self-interactions cause energy transfer from the hotter outer halo to the colder central region, thereby forming a core.  At the same time, collisional stripping of dwarf subhalos within the hotter MW host halo can deplete the abundances of satellites.  Early simulations, which focused primarily on the case of a constant (velocity-independent) scattering cross section, found that $\sigma/m_X \sim 1 - 10 \, \cmg$ flattened the central densities in dwarf galaxies in accordance with observations and $\sigma/m_X \sim 10 \, \cmg$ reduced significantly the number of MW subhalos~\cite{Dave:2000ar}.  

Subsequent studies, however, found rather serious problems with self-interacting DM due to conflicts with other observations.  The simulation of Ref.~\cite{Yoshida:2000uw} concluded that $\sigma/m_X \lesssim 0.1 \, \cmg$ is required to avoid core formation in cluster halos in conflict with gravitational lensing observations of CL 0024+1654.  Ref.~\cite{Miralda-Escude} argued that $\sigma/m_X \lesssim 0.02 \, \cmg$ is required by cluster ellipticity constraints, while Ref.~\cite{Gnedin:2000ea} showed that $\sigma/m_X \sim 0.3 - 10^4 \, \cmg$ is excluded by requiring that elliptical galaxy halos do not evaporate within hot cluster halos. Lastly, Ref.~\cite{Randall:2007ph} obtained $\sigma/m_X \lesssim 1 \, \cmg$ from the X-ray and lensing observations of the Bullet cluster.

More recently, there have been two major developments leading to a revival of self-interacting DM.  First, DM self-interactions need not have a cross section that is constant in velocity~\cite{Feng:2009mn,Ackerman:2008gi,Feng:2009hw,Ibe:2009mk,Loeb:2010gj,Tulin:2012wi}.  For light dark force mediators, once the momentum transfer becomes comparable to the mediator mass, the cross section begins to decrease rapidly (analogous to Rutherford scattering).  Since larger halos have larger characteristic velocities, the cross section can be large in dwarf galaxies ($v \sim 10 \, \kms$), but negligible on cluster scales ($v \sim 1000 \, \kms$) to evade the aforementioned constraints.  

Second, considerable progress has been made in numerical simulations of self-interacting DM~\cite{Vogelsberger:2012ku,Rocha:2012jg,Peter:2012jh,Zavala:2012us}.  In particular, the issue of self-interacting DM constraints from galaxy clusters was recently revisited in Refs.~\cite{Rocha:2012jg,Peter:2012jh}.  In these simulations, a very different conclusion was reached from earlier simulations.  In particular, the constraints from cluster halo triaxiality were found to be much weaker than previously estimated.  They conclude that previous works did not take into account that the observed ellipticity has contributions from regions well outside the core, and this region retains its triaxiality.  They also find that the residual triaxiality is larger than previously estimated \cite{Dave:2000ar}, and that the remaining discrepancy can be accounted for in the ellipticity scatter between different DM halos.  Furthermore, the authors also find that the tendency of subhalos to evaporate is not significant for $\sigma/m \sim 1\mbox{ cm}^2/\mbox{g}$.  Lastly, the cluster CK 0024+1654 used by Ref.~\cite{Yoshida:2000uw} is now known to be undergoing a merger along the line of sight, making it less useful as a comparison case with non-merging simulation data.

Overall, while the situation for self-interacting DM is not yet resolved, much progress has been made.  The most recent simulations have shown that $\sigma/m_X \sim 0.1 - 10 \, \cmg$ on dwarf scales is sufficient\footnote{Ref.~\cite{Zavala:2012us} found that $\sigma/m_X = 0.1 \, \cmg$ is too small, although the precise lower bound is unknown.} to solve the core-vs-cusp and too-big-to-fail problems~\cite{Vogelsberger:2012ku,Rocha:2012jg,Peter:2012jh,Zavala:2012us}, while constraints on MW and cluster scales require $\sigma/m_X \lesssim 0.1 - 1 \, \cmg$~\cite{Vogelsberger:2012ku,Rocha:2012jg,Peter:2012jh}.  It appears that all the data may be accounted for with a constant scattering cross section around $\sigma/m_X \sim 0.5 \, \cmg$.  On the other hand, particle physics models of self-interacting DM generically predict a velocity-dependent scattering cross section over a wide range of parameter space, as we discuss below.

%%%%%%%%%%%%%%%%%%%%%%%%%%%%%%%%%%%%%%%%%%%555
\section{Dark Forces and Dark Matter Scattering}
%%%%%%%%%%%%%%%%%%%%%%%%%%%%%%%%%%%%%%%%%%%%%%%5

In order to explain astrophysical observations on dwarf galaxy scales, the DM elastic scattering cross section must be 
\be \label{desiredxsec}
\sigma \sim 1 \; {\rm cm}^2 \, (m_X/{\rm g}) \approx 2 \times 10^{-24} \; {\rm cm}^2 \, (m_X/{\rm GeV}) \, ,
\ee
which is much larger than a typical weak-scale cross section $\sigma \sim 10^{-36} \; {\rm cm}^2$.  Therefore, this suggests the existence of a dark force boson $\phi$ that is much lighter than the weak scale.  

In this work, we consider a phenomenological approach where nonrelativistic DM scattering is described by a Yukawa potential
\beq
V(r) = \pm \frac{\alpha_X}{r} e^{-m_\phi r} \, , \label{potential}
\eeq
which can be either repulsive ($+$) or attractive ($-$).  This interaction arises for $\phi$ as a vector or scalar mediator, with interaction
\beq
\mathscr{L}_{\rm int} = \left\{ \begin{array} {ll} g_X \bar X \gamma^\mu X \phi_\mu & {\rm vector \; mediator} \\
g_X \bar X X \phi & {\rm scalar\; mediator} \end{array} \right.
\eeq
and dark fine structure constant $\alpha_X = g_X^2/(4\pi)$.  Scalar interactions are purely attractive, while a vector interaction is both attractive ($X \bar X$ scattering) and repulsive ($X X$ or $\bar X \bar X$ scattering).  Thus, in the vector case, asymmetric DM ($X$ only) will have purely repulsive interactions, while symmetric DM (equal $X,\bar X$) will have both attractive and repulsive interactions, with the total effective cross section given by the average of the two.

Numerical N-body simulations have investigated the impact of DM self-interactions on structure formation.  The relevant input is the differential cross section $d \sigma/ d \Omega$, as a function of the DM relative velocity $v$.  Since simulations track particle trajectories before and after collisions, the angular distribution over the scattering angle $\theta$ is important. However, to compare across different parameter regions, with different angular dependencies, it is useful to consider an integrated cross section that captures the relevant physics.  The usual quantity is the standard cross section $\sigma = \int d\Omega (d\sigma/d\Omega)$.  However, for light mediators, $\sigma$ receives a strong enhancement in the forward-scattering limit ($\cos\theta \to 1$), and for the purposes of affecting the DM distribution this enhancement is spurious since the DM particle trajectories are unchanged.  In the plasma literature, two additional cross sections are defined to parametrize transport~\cite{krstic:1999}, the transfer cross section $\sigma_T$ and the viscosity (or conductivity) cross section $\sigma_V$:
\beq
\sigma_T =  \int d\Omega \, (1-\cos\theta) \,\frac{d\sigma}{d\Omega}  \, , \qquad  \label{sigmaT}
\sigma_V =  \int d \Omega \, \sin^2 \theta \, \frac{d\sigma}{d\Omega} \, .
\eeq
The transfer cross section is weighted by $(1-\cos\theta)$, the fractional longitudinal momentum transfer, while the viscosity cross section is weighted by the energy transfer in the transverse direction, $\sin^2 \theta$.  The transfer cross section has been used in the DM literature to regulate the forward-scattering divergence.  On the other hand, the viscosity cross section weighs forward and backward scattering evenly.  It takes into account that forward and backward scattering affect the DM halo equally, since DM particles simply exchange trajectories that they would have had in the absence of a collision.   It also takes into account that we expect that perpendicular scattering is most efficient for ``thermalizing'' the DM halo and affecting structure observables.  

In addition, the transfer cross section obviously fails if DM scattering occurs between identical particles.  Taking quantum indistinguishability into account, both forward and backward scattering diverges, corresponding to poles in the $t$- and $u$-channel diagrams.  $\sigma_T$ regulates only the forward divergence, making it an inadequate description for the case of quantum indistinguishable particles.  Since both forward and backward scattering leave the DM distribution unchanged, the relevant cross section should regulate both divergences, which $\sigma_V$ does, but $\sigma_T$ does not.

In order to make contact with previous work, however, we focus on $\sigma_T$, rather than $\sigma_V$.  Under the assumption of classical distinguishibility in scattering, we find that $\sigma_T$ and $\sigma_V$ differ by less than a factor of two, with $\sigma_V$ for distinguishable and indistinguishable particles differing by another ${\cal O}(1)$ number.  Thus the overall effect both of distinguishability and of the transfer versus viscosity cross section is ${\cal O}(1)$.  For the purpose of presenting our results, we assume classical distinguishability and take $\sigma_T$ as a suitable measure for the effects of DM scattering on halo shapes.  Of course, a full-scale N-body simulation should make use of the angular information in the differential scattering cross section, $d \sigma / d \Omega$, and do away with the proxy of a transfer or viscosity cross section altogether, though in most cases the difference between the results using $\sigma_V$ or $\sigma_T$ versus $d\sigma/d\Omega$ will be small.  In Sec.~\ref{angular} below, we discuss the angular dependence in more detail and present benchmarks for simulation. 

The transfer cross section, computed perturbatively in $\alpha_X$ from Eq.~\eqref{potential}, is given by
\beq
\sigma_T^{\rm Born} = \frac{8\pi \alpha_X^2}{m_X^2 v^4} \Big( \log\big(1+m_X^2 v^2/m_\phi^2\big) -\frac{m_X^2 v^2}{m_\phi^2 + m_X^2 v^2} \Big) \label{born} \; .
\eeq 
for both attractive and repulsive potentials~\cite{Feng:2009hw}, where $v$ is the relative velocity.  This perturbative expression is valid only within the Born approximation, requiring $\alpha_X m_X/m_\phi \ll 1$.  Outside this limit, the Born approximation is not valid and non-perturbative corrections become crucially important.  

Within the non-perturbative regime, analytic formulae for $\sigma_T$ have been obtained only within the classical limit ($m_X v/m_\phi \gg 1$)~\cite{Feng:2009hw,Khrapak:2003,Khrapak:2004}, given for an attractive potential by
\beq
\sigma_T^{\rm clas} = 
\left\{\begin{array}{lc}
\frac{4 \pi}{m_\phi^2} \beta^2 \ln\left(1+\beta^{-1}\right) & \beta \lesssim 10^{-1} \\
\frac{8 \pi}{m_\phi^2} \beta^2 / \left(1+1.5 \beta^{1.65}\right) & \; 10^{-1} \lesssim \beta \lesssim 10^3 \\
\frac{\pi}{m_\phi^2} \left(\ln \beta+1-\frac{1}{2} \ln^{-1}\beta \right)^2 & \beta \gtrsim 10^3
\end{array} \right. \label{plasma} \, ,
\eeq
where $\beta \equiv 2 \alpha_X m_\phi/(m_X v^2)$.  Many previous works~\cite{Feng:2009hw,Ibe:2009mk,Loeb:2010gj,Aarssen:2012fx}, including recent N-body simulations~\cite{Vogelsberger:2012ku}, have focused specifically on the case where DM scattering is described by an attractive, classical cross section, given by Eq.~\eqref{plasma}.  We emphasize, however, that this case is just one out of many possibilities, and in general the non-perturbative regime remains largely unexplored.  We collect, for reference, the analytic formulae in the Appendix for the Born, attractive and repulsive classical, and $s$-wave resonance cases.

For a large parametric range of interest for DM self-interactions, {\it both} quantum mechanical and non-perturbative effects become important, and neither the Born nor classical approximations are valid.  The onset of these effects is governed by the conditions $\alpha_X m_X/m_\phi \gtrsim 1$ and $m_X v/m_\phi \lesssim 1$, respectively.  We denote this region of parameter space as the ``resonant regime,'' since one important effect is the appearance of quantum mechanical resonances in $\sigma_T$ corresponding to \mbox{(quasi-)bound} states in the potential.  

Within the resonant regime, there exists no analytic formula for $\sigma_T$, and it must be computed by solving the Schr\"{o}dinger equation directly using a partial wave analysis.  The differential scattering cross section is given by
\beq \label{diffsigma}
\frac{d \sigma}{d\Omega} = \frac{1}{k^2} \Big| \sum_{\ell = 0}^\infty (2 \ell + 1) e^{i \delta_\ell} P_\ell(\cos\theta) \sin \delta_\ell \Big|^2 \, ,
\eeq
where $\delta_\ell$ is the phase shift for a partial wave $\ell$.  In terms of the phase shifts, the transfer cross section is given by
\be
\frac{\sigma_T k^2}{4\pi} =  \sum_{\ell = 0}^{\infty} (\ell + 1) \sin^2 (\delta_{\ell+1} - \delta_\ell)  \, . \label{sigmaTsum}
\ee
To obtain $\delta_\ell$, one must solve the Schr\"{o}dinger equation for the radial wavefunction $R_\ell(r)$ for the reduced DM two-particle system, given by
\beq
\frac{1}{r^2} \frac{d}{dr} \Big( r^2 \frac{d R_{\ell}}{dr} \Big) + \Big( k^2 - \frac{\ell (\ell + 1)}{r^2} - 2\mu V(r) \Big) R_\ell = 0 \label{radial}
\eeq
with reduced mass $\mu = m_X/2$ and momentum $k = \mu v$.  The phase shift $\delta_\ell$ parametrizes the asymptotic solution for $R_\ell(r)$, given by
\beq \label{Rasymp}
\lim_{r \to \infty} R_\ell(r) \propto   \cos\delta_\ell \, j_\ell(kr) - \sin\delta_\ell \, n_\ell(kr)  \, ,
\eeq
where $j_\ell$ ($n_\ell$) is the spherical Bessel (Neumann) function.

%%%%%%%%%%%%%%%%%%%%%%%%%%%%%%%%%%%%%%%%%%%%%%5
\section{Numerical Scattering Results}
\label{sec:results}
%%%%%%%%%%%%%%%%%%%%%%%%%%%%%%%%%%%%%%%%%%%%%%%%%5

In this section, we present our numerical results.  First, we describe our numerical method for computing the DM self-interaction cross section $\sigma_T$.  Next, we investigate the velocity-dependence and angular-dependence of DM scattering.  For realistic particle physics models of self-interacting DM, scattering can possess a wide range of nontrivial dependence on velocity and scattering angle, whereas N-body simulations have considered isotropic scattering with constant or particular choices of velocity dependencies.

\subsection{Numerical Method}

To solve the Schr\"{o}dinger equation, it is useful to define the variables~\cite{Buckley:2009in}
\beq
\chi_\ell \equiv r R_\ell \, , \quad x \equiv \alpha_X m_X r  \, , \quad a \equiv \frac{v}{2\alpha_X} \, , \quad b \equiv \frac{\alpha_X m_X}{m_\phi} \; , \label{vardefs}
\eeq
such that Eq.~\eqref{radial} can be expressed as
\beq \label{radial2}
\left( \frac{d^2 }{d x^2} +  a^2 - \frac{\ell(\ell+1)}{x^2} \pm \frac{1}{x} \, e^{-x/b} \right) \chi_\ell(x) = 0 \; .
\eeq
To compute $\sigma_T$, we first compute $\delta_\ell$ for given $(a,b,\ell)$ as follows.
\begin{enumerate}
\item We impose an initial condition for $\chi_\ell$ and $\chi_\ell^\prime$ at a point $x=x_i$ close to the origin.  For \mbox{$x_i \ll b, (\ell+1)/a$}, Eq.~\eqref{radial2} is dominated by the angular momentum term, and we expect $\chi_\ell(x) \propto x^{\ell +1}$.  Thus, we take $\chi_\ell(x_i) = 1$ and $\chi_\ell^\prime(x_i) = (\ell+1)/x_i$; the overall normalization is irrelevant. 
\item We solve Eq.~\eqref{radial2} numerically within the domain $x_i \le x \le x_m$.  The matching point $x_m$ is determined by the condition $a^2 \gg \exp(-x_m/b)/x_m$, where the potential term is suppressed compared to the kinetic term.
\item At $x=x_m$, we match $\chi_\ell$ (and its first derivative) onto the asymptotic solution, given by
\be
\chi_\ell(x) \propto x \, e^{i \delta_\ell} \big(\cos\delta_\ell \, j_\ell(a x) - \sin \delta_\ell \, n_{\ell}(a x) \big) \; . \label{asympsol}
\ee
Inverting Eq.~\eqref{asympsol}, the phase shift is given by
\be
\tan \delta_\ell = \frac{a x_m \, j^\prime_\ell(a x_m) - \beta_\ell \, j_\ell(a x_m) }{a x_m \, n^\prime_\ell(a x_m) - \beta_\ell \, n_\ell(a x_m) } \; , \quad \beta_\ell = \frac{x_m \chi_\ell^\prime(x_m)}{\chi_\ell(x_m)} - 1
\ee
 in terms of our numerical solution for $\chi_\ell$ at $x_m$.  Our numerical method makes an initial guess for $(x_i,x_m)$ and computes $\delta_\ell$, and then successively decreases (increases) $x_i$ ($x_m$) until $\delta_\ell$ converges at $1\%$.
\item The last step is computing $\sigma_T$ by summing Eq.~\eqref{sigmaTsum} over $\ell$, truncating at $\ell_{\rm max}$.  We iterate $\ell_{\rm max}$ until $\sigma_T$ converges to $1\%$ and $\delta_{\ell_{\rm max}} < 0.01$ through ten successive iterations.  This condition is quite conservative, typically summing many more $\ell$-modes than required. 
\end{enumerate}
 For a given $(a,b)$, we can then express $\sigma_T$ in terms of the physical parameters $(m_X, m_\phi, \alpha_X, v)$.  Our numerical code for this solution was written using \texttt{Mathematica}.

\begin{figure*}[t]
\includegraphics[scale=0.64]{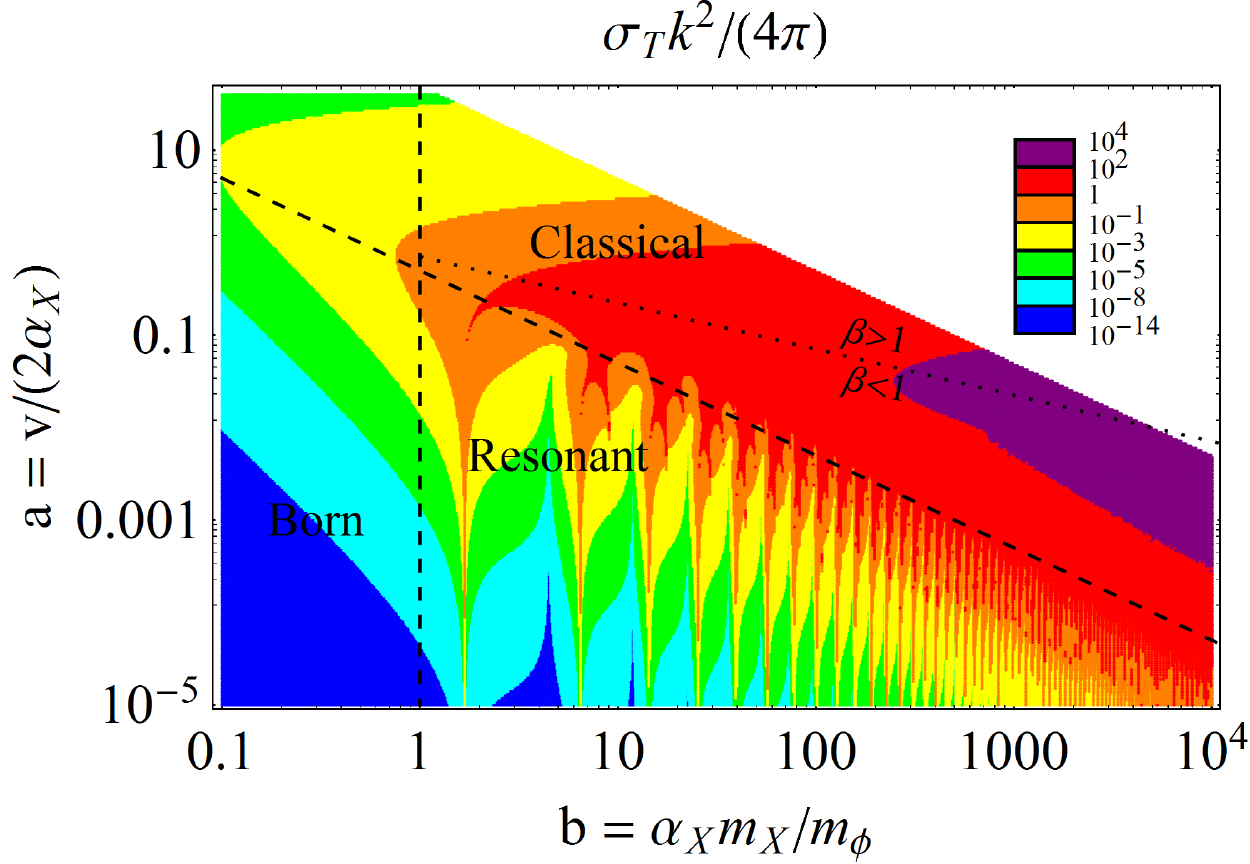}
\includegraphics[scale=0.64]{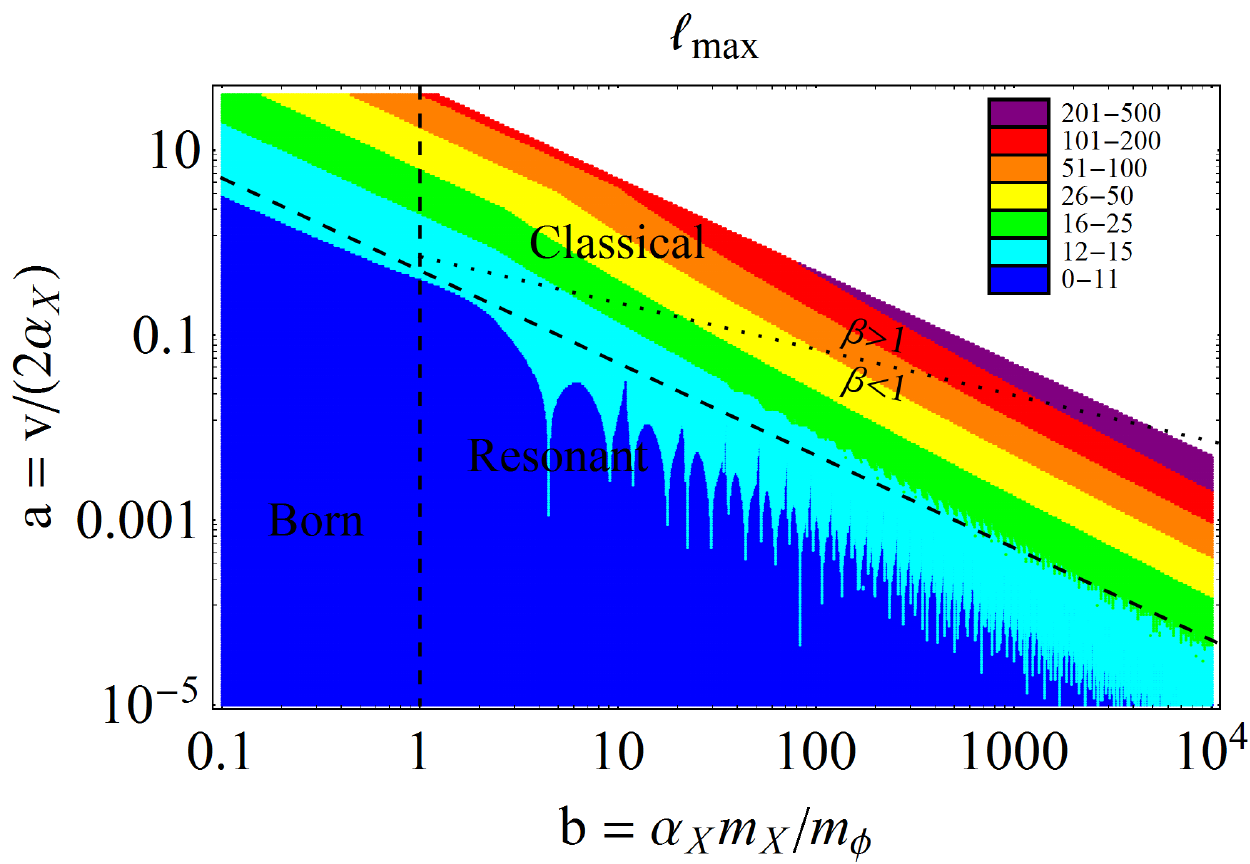}
\caption{Colored regions show parameter points $(a,b)$ within our numerical scan, with the corresponding values of $\sigma_T k^2/(4\pi)$ (left) and $\ell_{\rm max}$ (right) at each point.  The classical, Born, and resonant regimes are delineated by solid lines. \label{scan}}
\end{figure*}

With our numerical method in hand, we performed a fine-grained scan over $2\times 10^{5}$ parameter points $(a,b)$.  Fig.~\ref{scan} gives a birds-eye view of our full numerical dataset, with the colored points showing the parameters $(a,b)$ in our scan.  In the left panel, the different colors correspond to the computed value of $\sigma_T k^2/(4\pi)$ obtained from Eq.~\eqref{sigmaTsum}, with the corresponding value of $\ell_{\rm max}$ shown in the right panel.  The white region (upper right) was omitted from our scan.  The solid lines at $b=1$ and $2ab=1$ delineate the Born regime ($b \ll 1$), the classical regime ($2ab \gg 1$), and the resonant regime ($b \gtrsim 1$ and $2ab \lesssim 1$).  The latter exhibits a pattern of resonances in $\sigma_T$.

\begin{figure}
\includegraphics[scale=0.62]{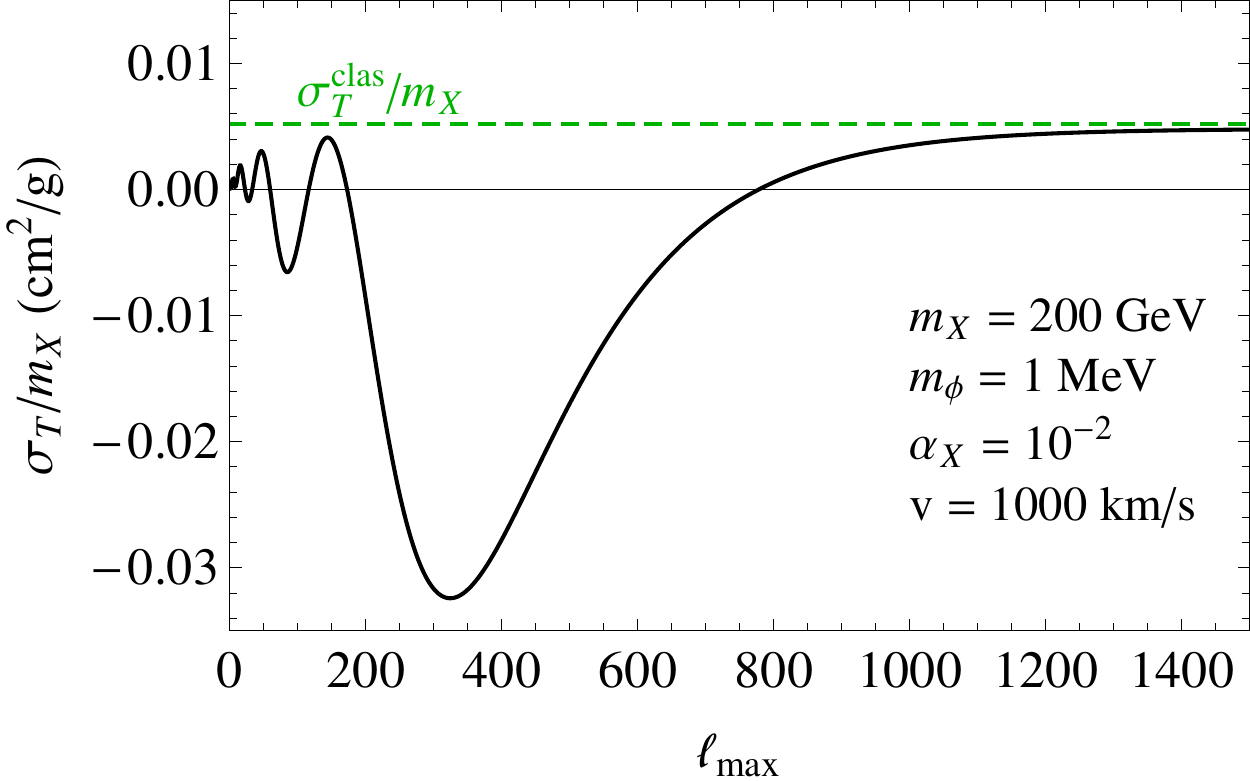}
\includegraphics[scale=0.92]{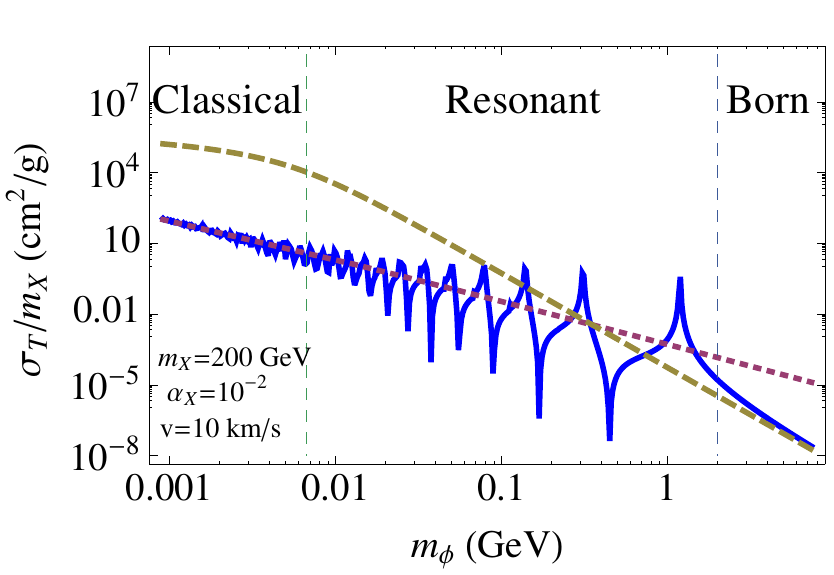}
\caption{{ Left}: Numerical calculation of $\sigma_T/m_X$, truncated at fixed $\ell_{\rm max}$, showing convergence with increasing $\ell_{\rm max}$. The parameter point chosen corresponds to the classical regime with an attractive potential.  The convergence to the classical analytic result shown by dashed line.  {Right}: Numerical calculation (solid blue) of $\sigma_T/m_X$ versus $m_\phi$, showing convergence to the classical analytical formula (dotted pink) and Born approximation (dashed gold) in the classical and Born regimes.}
\label{Lmaxconv}
\end{figure}

There is an importance difference between our method and that of Ref.~\cite{Buckley:2009in}, which performed a similar calculation of $\sigma_T$ within the resonant regime, albeit for a limited choice of parameters.  Ref.~\cite{Buckley:2009in} obtained $\delta_\ell$ by matching onto an asymptotic form $\chi_\ell(x) \propto \sin( a x - \pi \ell/2 + \delta_\ell)$, which is equivalent to Eq.~\eqref{asympsol} for $x\to \infty$.  For finite $x$, this form is valid only if {\it both} the Yukawa and angular momentum terms in Eq.~\eqref{radial2} are suppressed compared to the kinetic term, whereas Eq.~\eqref{asympsol} requires {\it only} the Yukawa term to be suppressed.  Therefore, as $\ell$ is increased, the method of Ref.~\cite{Buckley:2009in} requires integrating Eq.~\eqref{radial2} to much larger $x$ than in our method, and is therefore much less efficient.  Thus, Ref.~\cite{Buckley:2009in} truncates at $\ell_{\rm max} = 5$ in their calculation, whereas we are able to perform efficient calculations with $\ell_{\rm max} \sim 1000$.  We demonstrate this point in Fig.~\ref{Lmaxconv}, showing how $\sigma_T$ depends on $\ell_{\rm max}$ for one parameter choice in the classical regime.  Our numerical calculation (solid line) converges for $\ell_{\rm max} \gtrsim 1000$, in good agreement with the classical cross section (dashed line).\footnote{The reader should not be troubled by the fact that $\sigma_T$ can be negative for certain values of $\ell_{\rm max}$.  Due to the fact that the momentum and orbital angular momentum operators do not commute, the transfer cross section, defined in terms of momentum eigenstates, is a physical quantity only in the limit $\ell_{\rm max} \to \infty$, not for a particular value of $\ell$.}  

We can also see the convergence to classical and Born analytic formulae in the right panel of Fig.~\ref{Lmaxconv}. The dashed gold and dotted pink lines show the results for the Born and classical analytic formulae, and we see that in the regime of validity, our numerical results (solid blue line) agree well with the analytic formulae. In the quantum resonant regime, neither of the analytic formulae reproduce the behavior of the resonant peaks and anti-resonant valleys. Also note that the Born approximation over-estimates the cross section in the classical regime.

%%%%%%%%%%%%%%%%%%%%%%%%%%%%%%%%%%%%%%%
\subsection{Velocity-dependence in dark matter scattering}
%%%%%%%%%%%%%%%%%%%%%%%%%%%%%%%%%%555

The most important feature that emerges from our numerical study is the highly nontrivial velocity-dependence of $\sigma_T$ within the resonant regime.  While previous studies have focused on either constant $\sigma_T$ or specific $v$-dependencies, a rich array of possibilities can arise in general, and the velocity behavior can be rather complicated.  

\begin{figure}
\includegraphics[scale=0.64]{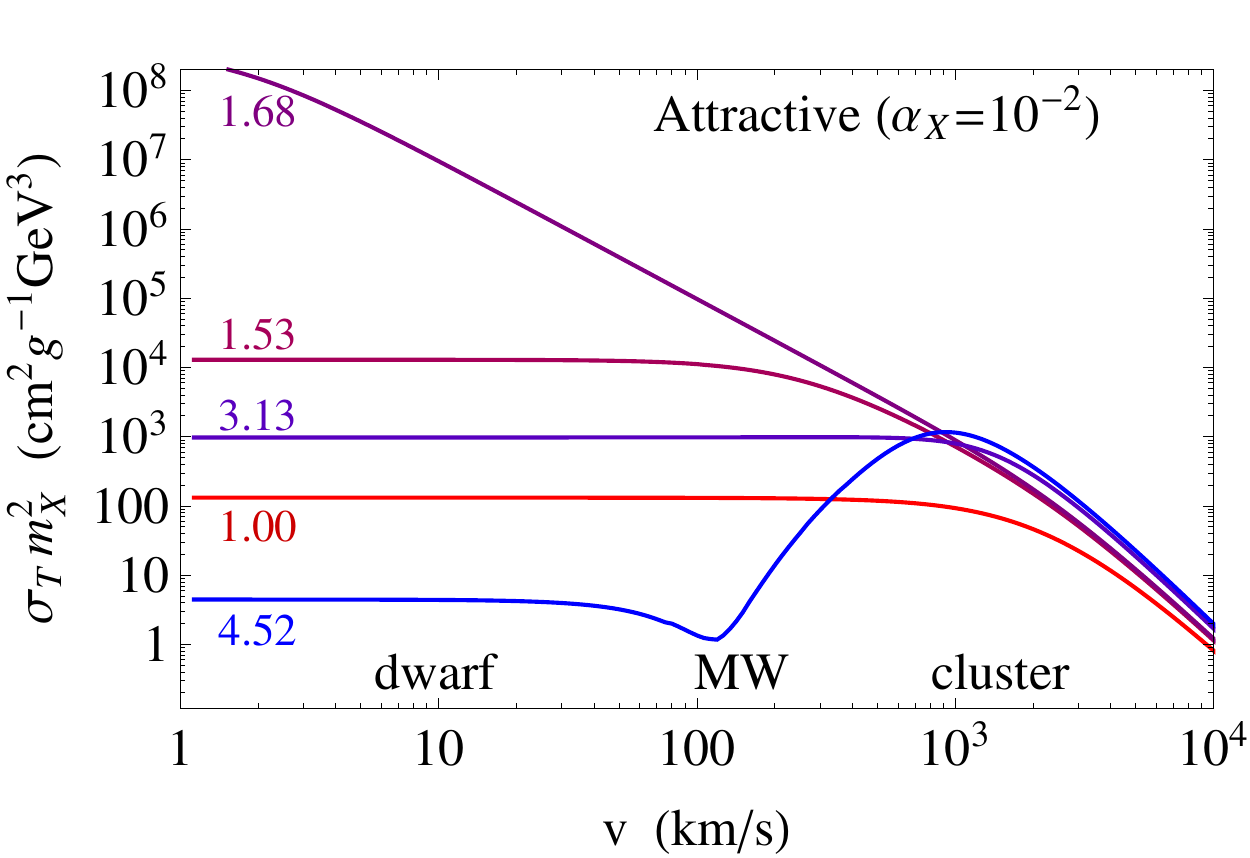} \includegraphics[scale=0.64]{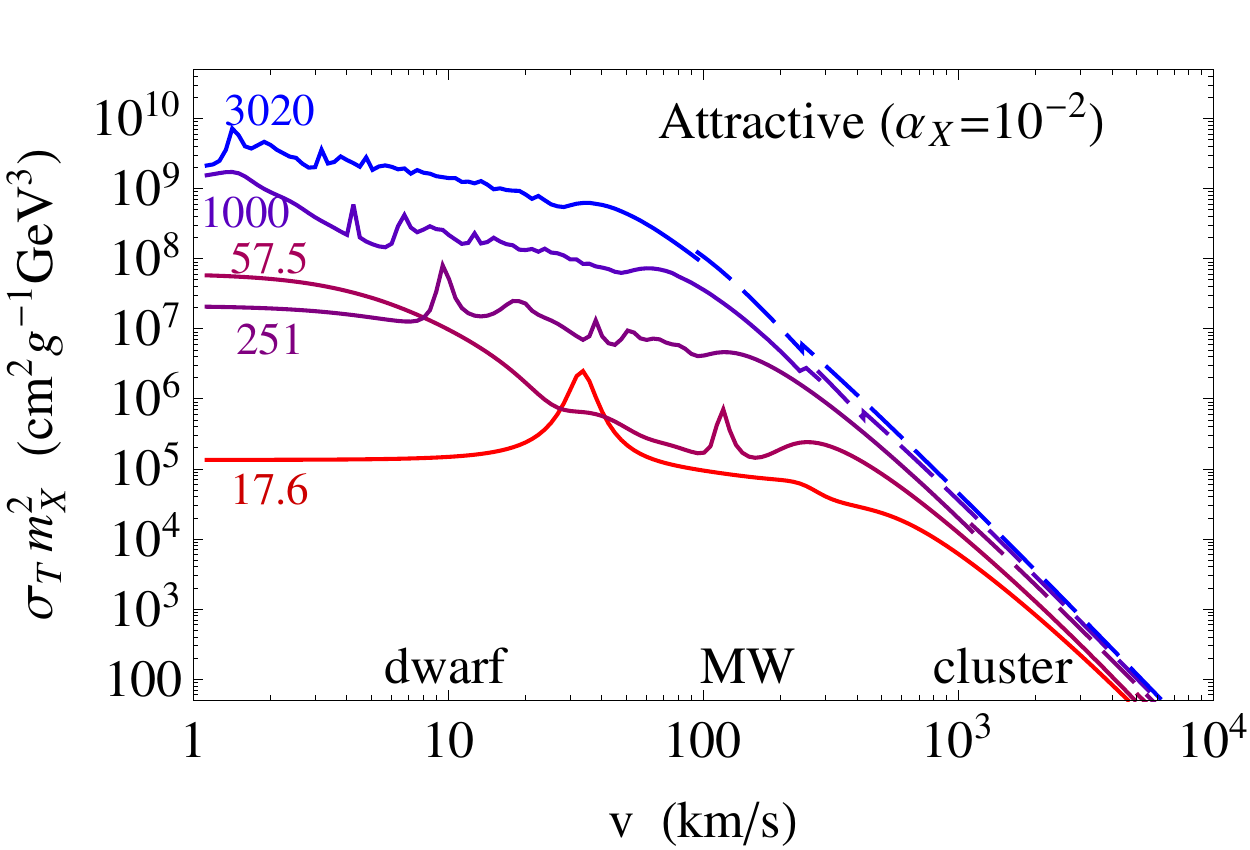}
\caption{Velocity-dependence of the scattering cross section for an attractive potential with $\alpha_X=10^{-2}$, computed numerically (solid lines), for various values of $b$ labeling each curve.  Numerical values indicate $\alpha_X m_X/m_\phi$ chosen for each curve.  Dashed lines show extrapolation using classical formulae.  Each curve can be normalized to $\sigma_T/m_X \sim 1 \; \cmg$ on dwarf halo scales ($v\approx 10 \;\kms$) by dividing by $m_X^3$ for a particular choice of $m_X$. }
\label{attxsec}
\end{figure}

\begin{figure}
\includegraphics[scale=0.64]{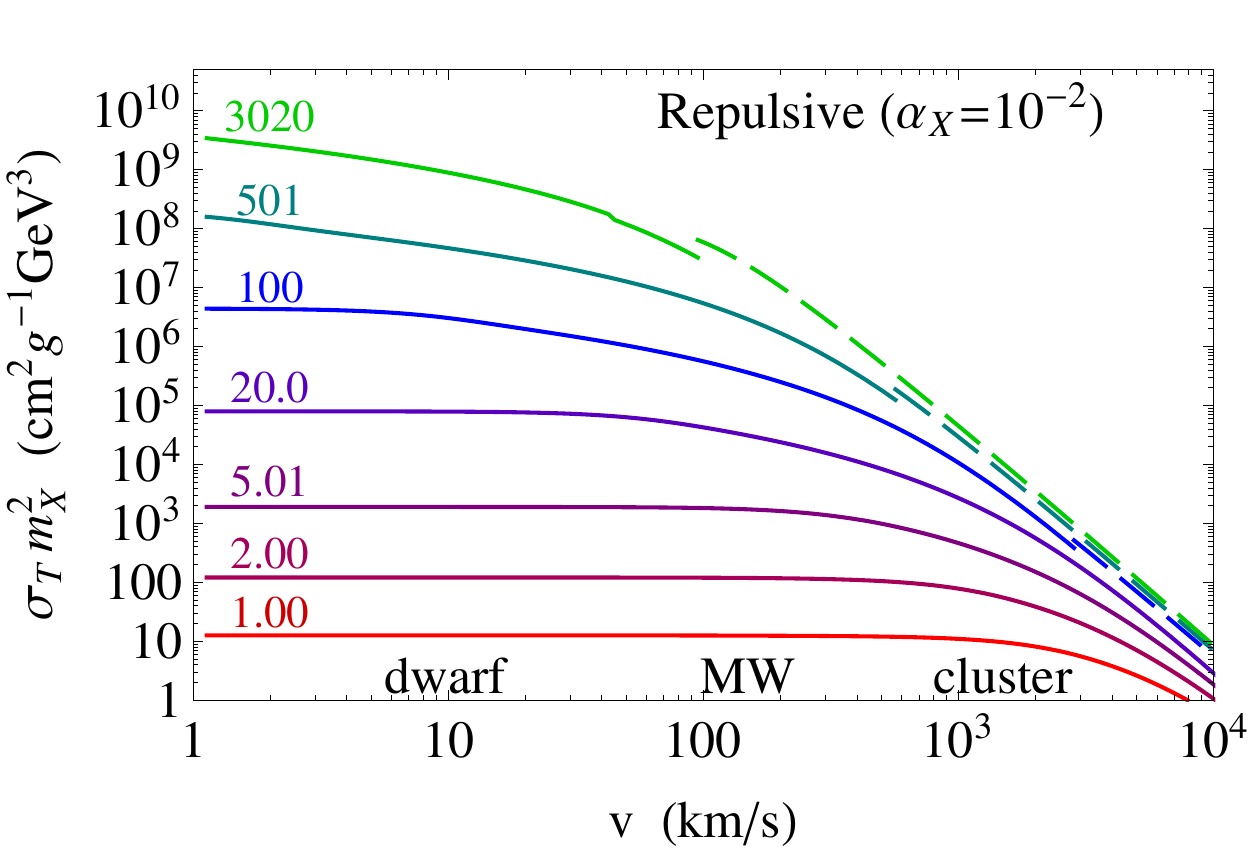}
\caption{Velocity-dependence of the scattering cross section, as Fig.~\ref{attxsec}, but for a repulsive potential.}
\label{repxsec}
\end{figure}

In Fig.~\ref{attxsec}, we show the cross section as a function of velocity for an attractive potential with $\alpha_X = 10^{-2}$.  Each curve corresponds to a different value for $b$ (where $b\equiv \alpha_X m_X/m_\phi$), as indicated by the numerical values in the figures.  The quantity $\sigma_T m_X^2$ is a useful normalization for the cross section since, for fixed $\alpha_X$, it depends on $v$ and $m_X/m_\phi$ only (as opposed to $m_X$ and $m_\phi$ separately).  Thus, to obtain the required level scattering in dwarf halos, each curve can be normalized to $\sigma_T/m_X \sim 1 \; \cmg$ at $v \approx 10$ km/s by choosing $m_X$ appropriately, which also fixes $m_\phi$.\footnote{Although the cross sections shown in Figs.~\ref{attxsec} and \ref{repxsec} have fixed $\alpha_X=10^{-2}$, these results can be generalized to other values of $\alpha_X$ by a shift in the horizontal and vertical axes.  Effectively, this shift amounts to relabeling the axes by $v \to v \times (10^{-2} /\alpha_X)$ and $\sigma_T m_X^2 \to \sigma_T m_X^2 \times (\alpha_X/10^{-2})^2$.}

The cross sections shown in Fig.~\ref{attxsec} exhibit a wide variety of behaviors and features.  The sequence of different cross sections, ascending from $b=1$ to $b \sim 1000$, illustrate the onset of resonance features beyond the Born regime ($b \ll 1$).  Increasing $b$, we first see the appearance of an $s$-wave resonance for $b = 1.68$; the phase shift behaves as $|\delta_0| \to \pi/2$ for $v \to 0$, and so the cross section becomes strongly enhanced, growing as $\sigma_T \to 16\pi/(m_X^2 v^2)$ on-resonance.  Moving to larger values of $b$, the cross section becomes reduced, and we see the appearance of an $s$-wave  antiresonance for $b=4.52$, where the cross section is strongly suppressed at low velocity.  Higher $\ell$-mode resonances appear as peaks at finite $v$ where $\sigma_T$ is enhanced.  For $b=17.6$ we note the appearance of a $p$-wave resonance at $v \approx 30 \; \kms$, and for higher $b$, spectral features become increasingly prevalent.  At high velocity, all cross sections converge to the same Coulomb result, $\sigma_T m_X^2 \propto v^{-4}$, independent of $m_\phi/m_X$.

In Fig.~\ref{repxsec}, we show a similar set of results for the cross section arising from a repulsive interaction, with $\alpha_X = 10^{-2}$.  Unlike the attractive case, no resonances arise in the ``resonant'' regime, since there are no bound states in the potential.  However, the cross section exhibits a clear velocity dependence where scattering on dwarf scales can be enhanced compared to larger scales.  Larger values of $b$ (i.e., smaller $m_\phi$, for fixed $m_X$) correspond to a longer range force, enhancing $\sigma_T$.

%%%%%%%%%%%%%%%%%%%%%%%%%%%%%%%%
\subsection{Angular dependence in dark matter scattering}
%%%%%%%%%%%%%%%%%%%%%%%%%%%%%%%%
\label{angular}

Since N-body simulations track particle trajectories before and after collisions, it is required to know the differential cross section and its dependence on the scattering angle $\theta$.  Although for $s$-wave scattering, $d \sigma/d \Omega$ is isotropic, more complicated angular dependencies arise in a wide range of parameter space.  In general, if DM scattering is velocity-dependent, then often the differential cross section carries a nontrivial angular dependence.

First, we investigate the impact of anisotropic scattering within the classical regime.  The numerical simulation of Ref.~\cite{Vogelsberger:2012ku} considered a velocity-dependent cross section given by Eq.~\eqref{plasma}, corresponding to an attractive interaction.  Here, we consider one specific benchmark point from this work (denoted therein as ``RefP2''), shown to solve small scale structure anomalies.  This benchmark is parametrized phenomenologically by $\sigma_T^{\rm max}/m_X = 3.5 \, \cmg$ and $v_{\rm max} = 30 \, \kms$; these quantities are related to the underlying parameters $(m_X, m_\phi, \alpha_X)$ by $\sigma_T^{\rm max} \equiv 22.7/m_\phi^2$ and $v_{\rm max}^2 \equiv  2 m_\phi \alpha_X/(\pi m_X)$.\footnote{To clarify this notation, we note that the quantity $\sigma_T v$ is maximized at $v=v_{\rm max}$, at which $\sigma_T = \sigma_T^{\rm max}$~\cite{Loeb:2010gj}.}  We emphasize that Ref.~\cite{Vogelsberger:2012ku} assumed in their simulation an isotropic differential cross section given by $d \sigma/d \Omega = \sigma_T/(4\pi)$.  With our numerical solution in hand, we can check whether this approximation is justified.

\begin{figure}
\includegraphics[scale=0.6]{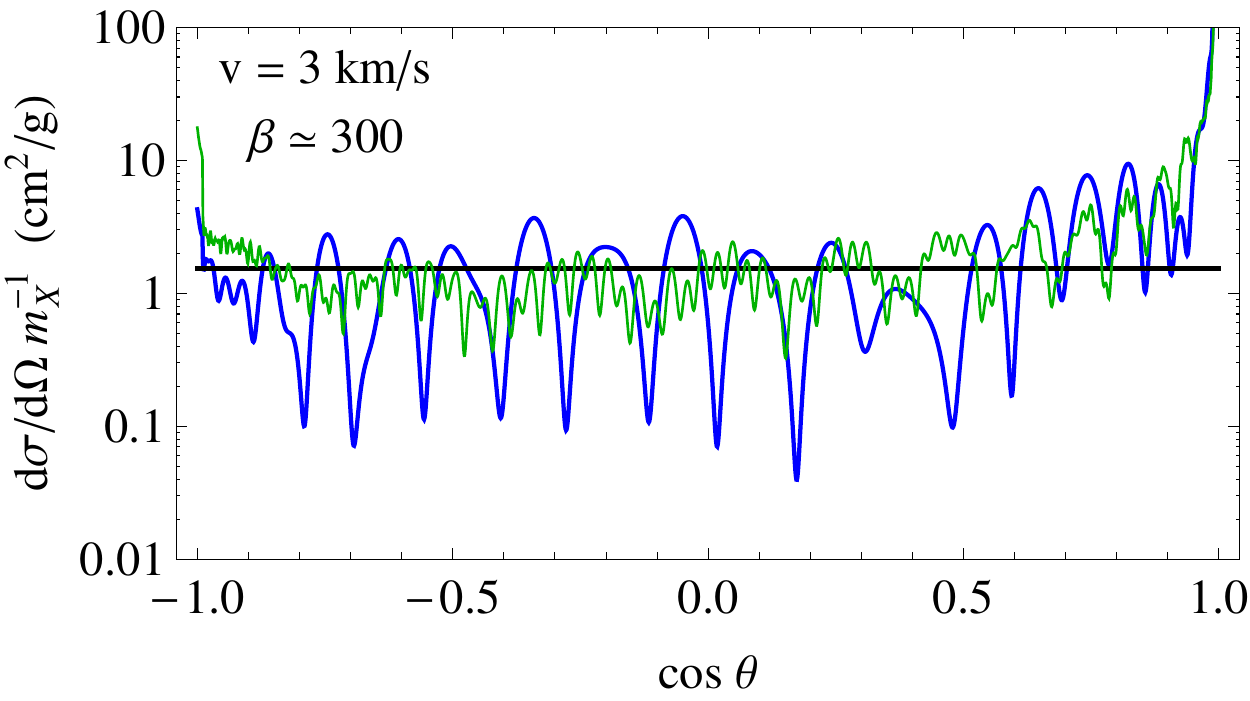} \includegraphics[scale=0.6]{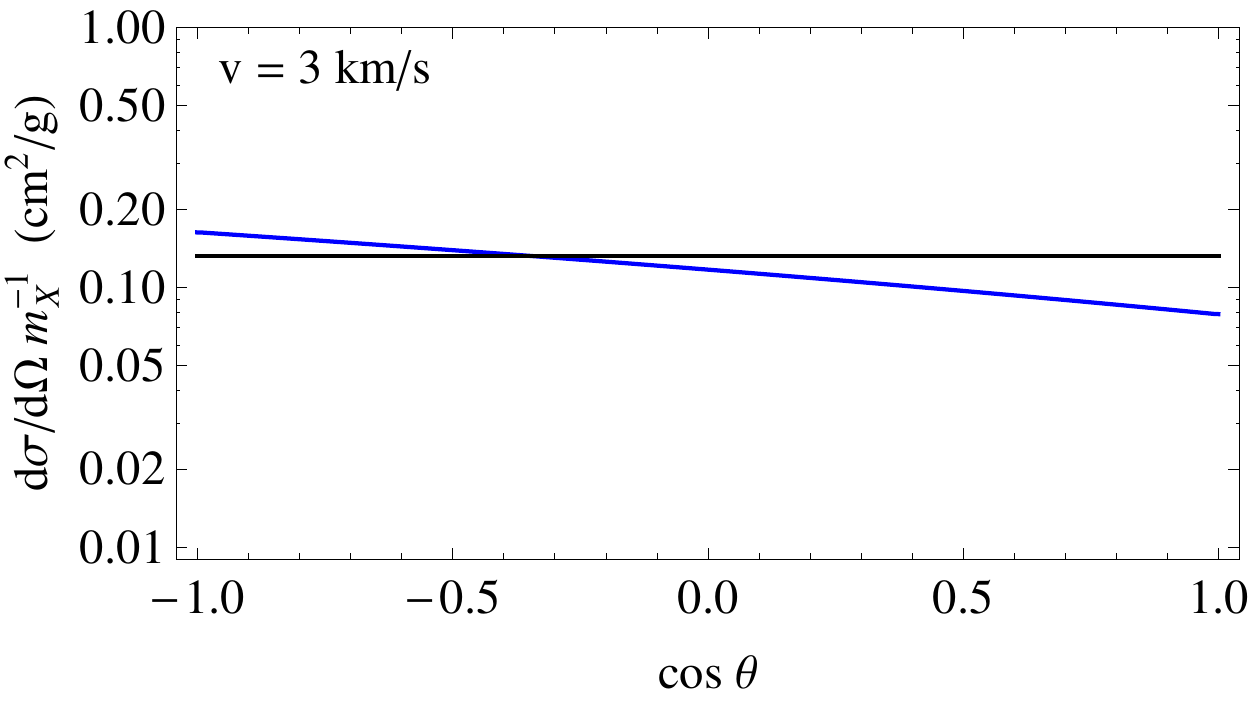}\\
\includegraphics[scale=0.6]{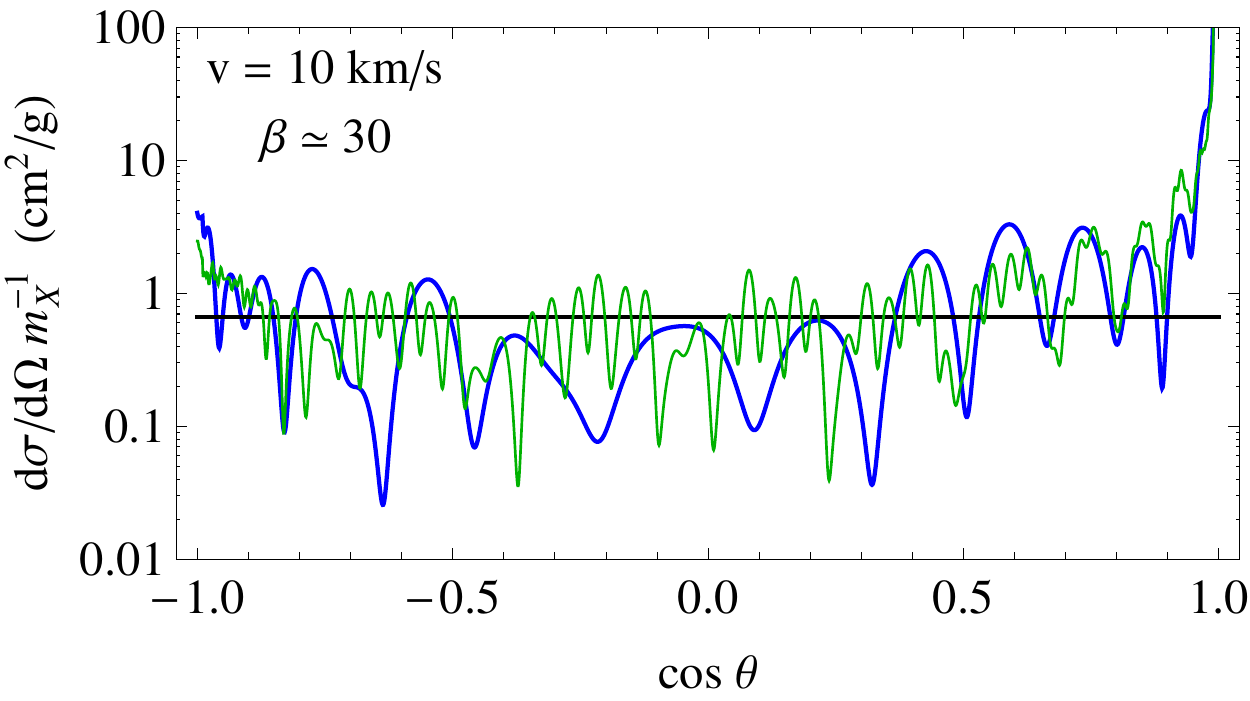} \includegraphics[scale=0.6]{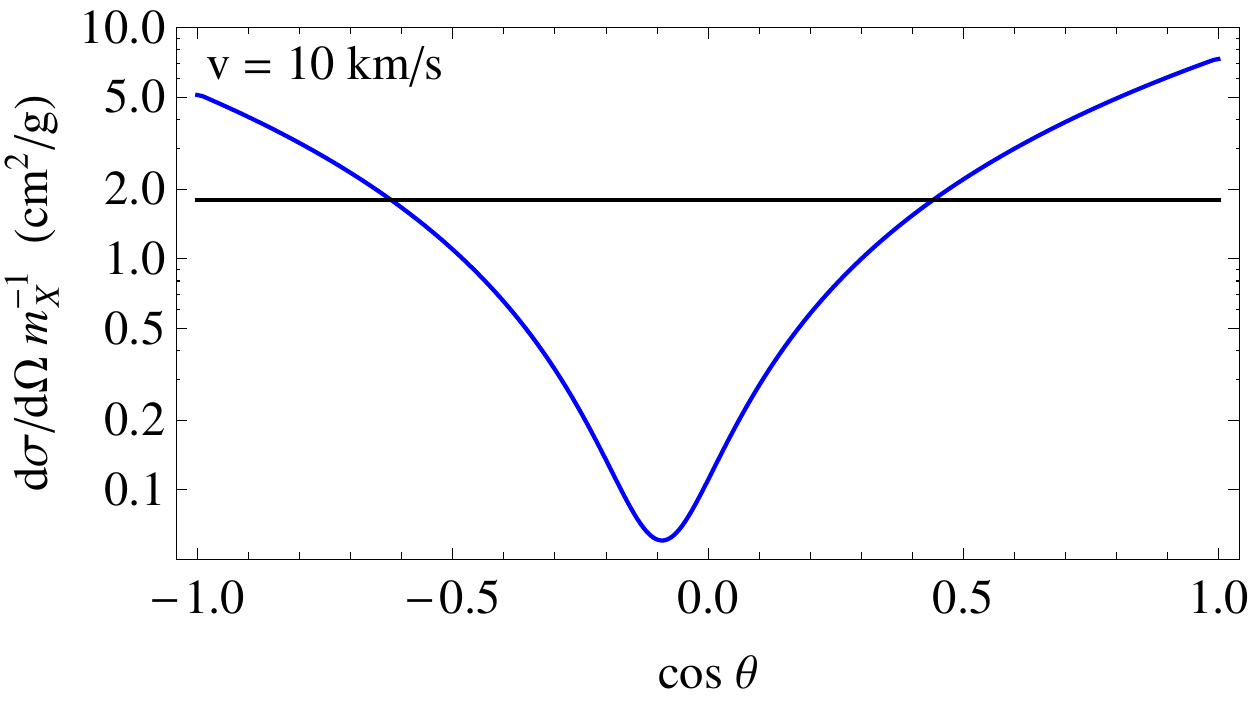} \\
\includegraphics[scale=0.6]{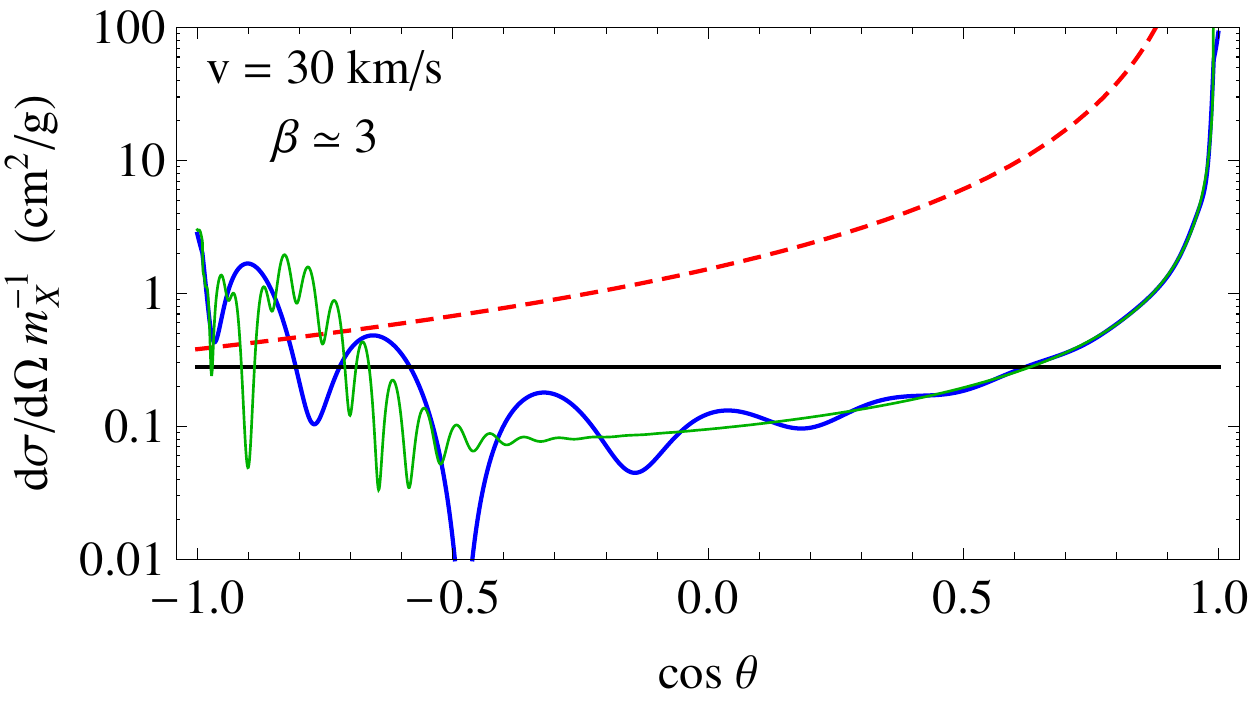} \includegraphics[scale=0.6]{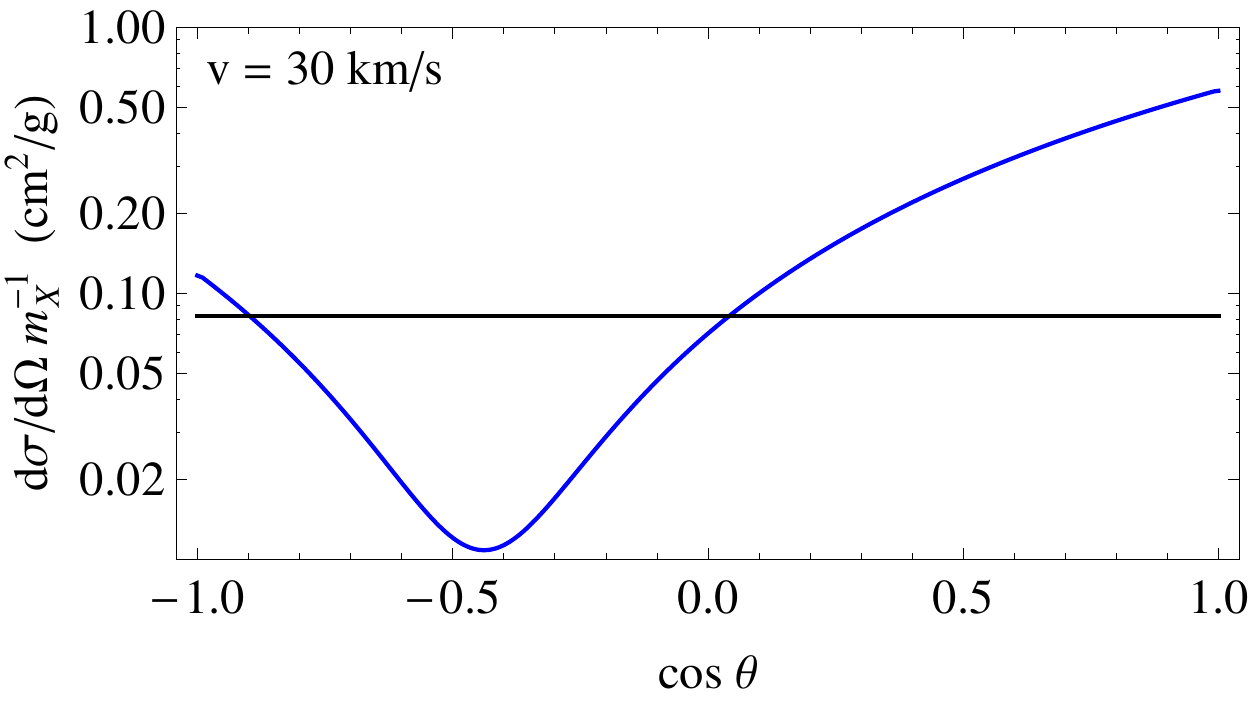} \\
\includegraphics[scale=0.6]{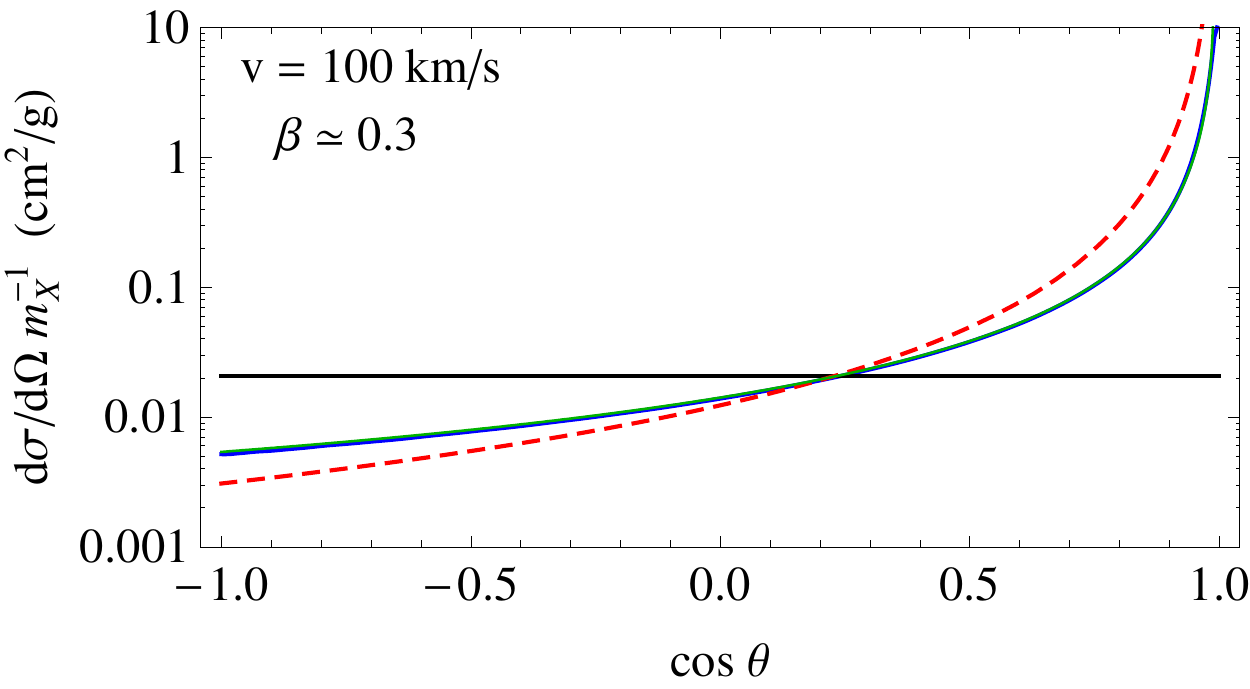} \includegraphics[scale=0.6]{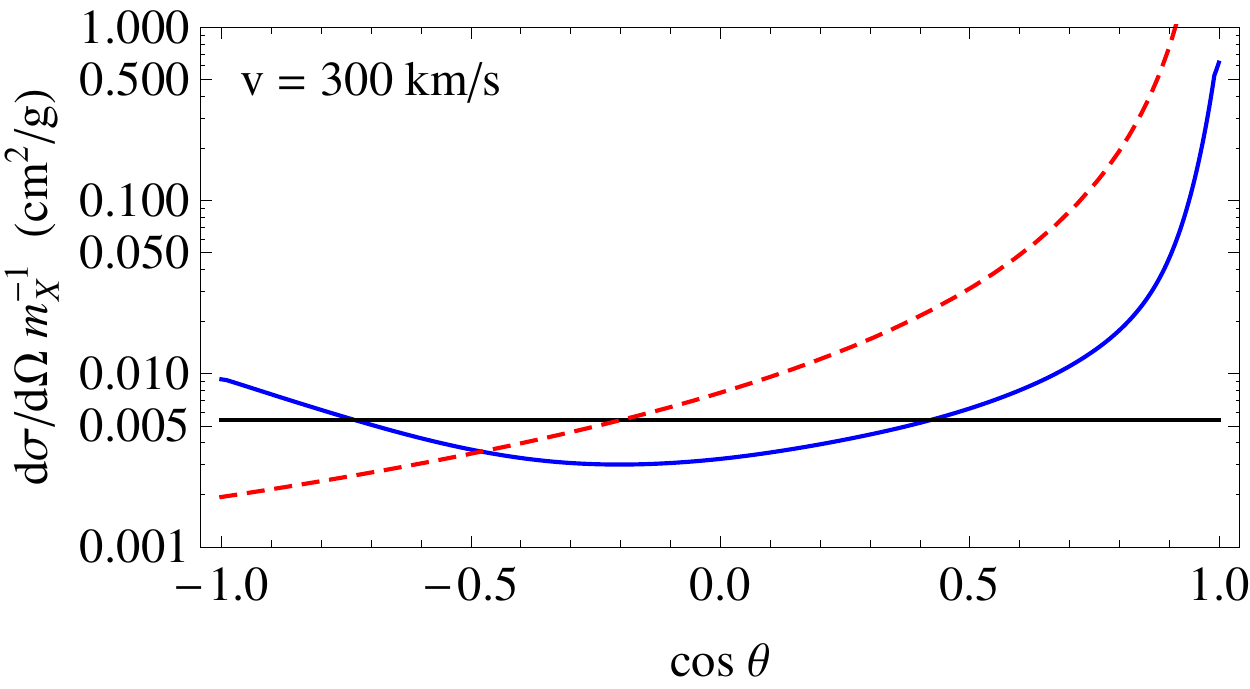} \\
\includegraphics[scale=0.6]{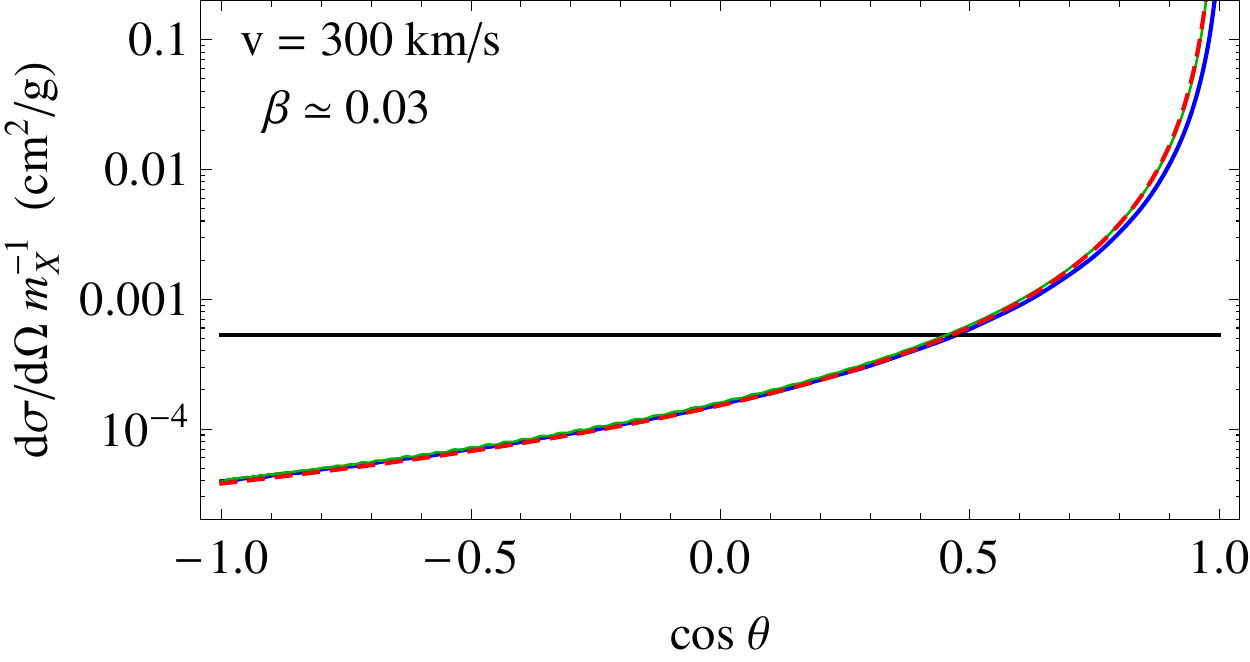} \includegraphics[scale=0.6]{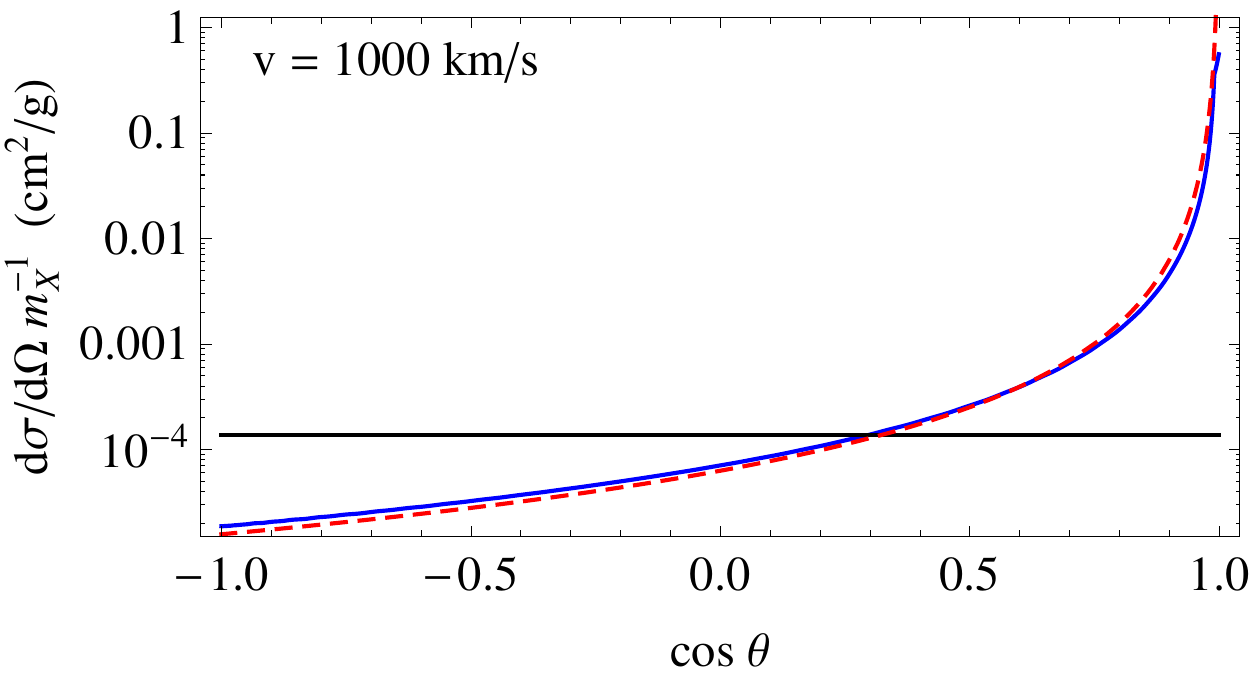} 
\caption{Left: numerical solution for $d\sigma/d\Omega \, m_X^{-1}$ for Ref.~\cite{Vogelsberger:2012ku} benchmark point for $m_X v/m_\phi = 10$ (thick blue) and $50$ (thin green).  Right: numerical solution for $d\sigma/d\Omega \, m_X^{-1}$ for benchmark point with $p$-wave resonance at $v \approx 10 \, \kms$ (solid blue).  Exact results are compared to the isotropic approximation $d \sigma/d \Omega = \sigma_T/(4\pi)$ (flat black) and the Rutherford formula (dashed red).}
\label{diffxsec}
\end{figure}

In Fig.~\ref{diffxsec} (left), we present our results for $d \sigma/ d\Omega$ for the RefP2 benchmark point, with each panel corresponding to a different velocity.  The horizontal black lines show the isotropic approximation $d \sigma/d \Omega = \sigma_T/(4\pi)$ adopted by Ref.~\cite{Vogelsberger:2012ku}.  The solid curves show our numerical calculation of $d \sigma/d \Omega$.  Although $v_{\rm max}$ and $\sigma_T^{\rm max}/m_X$ are fixed, an additional input is required to fix the three parameters $(m_X, m_\phi, \alpha_X)$.  We have taken $m_X v/m_\phi = 10$ (thick blue curve) and $m_X v/m_\phi = 50$ (thin green curve)\footnote{For visual clarity, we have smoothed these curves by averaging each point over an interval $\Delta \cos\theta = \pm 0.01$ to eliminate small-angle features.}; to the extent that these curves overlap, $d \sigma/d \Omega$ does not depend on this additional parametric freedom.  The dashed red line shows the usual Rutherford formula $d \sigma/ d\Omega = \alpha_X^2/( m_X^2 v^4 \sin^4 \theta/2)$.  From these plots, we conclude:
\begin{itemize}
\item At small velocity, $d\sigma/d\Omega$ has a nontrivial angular dependence, with many small-scale angular features oscillating about a nearly flat profile.  Since astrophysical structure observables are likely insensitive to small-angle features, we conclude that the isotropic approximation appears well-justified in this regime.  This behavior is expected since $\beta \gg 1$ corresponds to the strong coupling limit, and the Yukawa potential approaches the hard sphere limit with radius set by $m_\phi^{-1}$, with $d\sigma/d \Omega$ flat.
\item At large velocity, $d\sigma/d\Omega$ becomes peaked for forward-scattering ($\cos\theta \to 1$).  This behavior is expected since $\beta \ll 1$ corresponds to the Coulomb limit, and $d \sigma/d \Omega$ is well-approximated by the Rutherford formula.  We conclude that an isotropic approximation is not valid in this limit.  However, since the cross section is suppressed at larger velocity, this discrepency may not be important.
\end{itemize}
Similar conclusions apply to other parameter points in the classical regime: scattering is approximately isotropic for $\beta \gtrsim 1$, but becomes forward-peaked for $\beta \lesssim 1$.

Next, we consider a benchmark parameter point within the resonant regime: $m_X = 100$ GeV, $m_\phi = 17$ MeV, $\alpha_X = 3\times 10^{-3}$.  These parameters have been chosen to give a $p$-wave resonance on dwarf scales, with a peak at $v = 10 \, \kms$ with $\sigma_T/m_X = 22.5 \, \cmg$.  In Fig.~\ref{diffxsec} (right), we show our numerical results for $d \sigma/d \Omega$ (solid blue curves), with each panel corresponding to a different velocity, compared to the isotropic approximation $d \sigma/d \Omega = \sigma_T/(4\pi)$ (horizontal black lines).  At small $v$, scattering is predominantly $s$-wave, with $d \sigma/d\Omega$ nearly flat.  At $v = 10 \, \kms$, the $\ell=1$ term dominates, enhancing the scattering cross section and giving an angular dependence of $d \sigma/d \Omega \propto \cos^2\theta$.  For larger $v$, higher $\ell$ modes become important, and $d \sigma/d \Omega$ becomes forward-peaked, approaching the Rutherford formula.  For a $p$-wave resonance, it is clear that the angular dependence is crucial. Although $\sigma_T/m_X$ may be strongly enhanced on dwarf scales, the impact on astrophysical structure observables is likely less pronounced.  The $p$-wave angular distribution is weighted toward forward or backward scattering, whereas we expect structure observables to be more sensitive to perpendicular scattering ($\cos\theta \approx 0$).

\section{Parameter space for self-interacting dark matter}

In this section, we show how bounds from astrophysical observations of structure map onto the underlying DM particle physics parameter space.  Within our simple framework, there are only three parameters $(m_X, m_\phi, \alpha_X)$, as well as one overall sign corresponding to a repulsive or attractive potential in Eq.~\eqref{potential}.  For a given parameter choice, we compute $\sigma_T(v)$ either numerically or using the Born or classical analytic approximations (where valid).  However, the DM scattering probability within a halo is determined not by one fixed $v$, but rather by a convolution over different velocities and densities as a DM particle traverses the halo, requiring detailed N-body simulations which are beyond the scope of our work. Instead, we consider the velocity-averaged transfer cross section $\langle \sigma_T\rangle$ as a suitable proxy for the quantity being constrained by astrophysical bounds.  Averaging over the initial DM velocities $\vec{v}_{1,2}$ with exponential weight, we have
\be
\langle \sigma_T \rangle = \int \frac{ d^3 v_1 d^3 v_2}{(\pi v_0^2)^{3}} \, e^{-v_1^2/v_0^2}  \, e^{-v_2^2/v_0^2} \, \sigma_T(|\vec{v}_1 -\vec{v}_2|) = \int \frac{d^3 v}{(2\pi v_0^2)^{3/2}} \, e^{-\frac{1}{2} v^2/v_0^2} \, \sigma_T(v) \; ,
\ee
where $v_0$ is the most probable velocity and $v = |\vec{v}_1 -\vec{v}_2|$ is the relative velocity.  We choose $v_0$ to be characteristic of different size halos, described below.  Although velocity-averaging is clearly irrelevant for a constant cross section, it is especially important for strongly velocity-dependent cross sections ({\it e.g.}, resonant features).  

\begin{figure}
\includegraphics[scale=0.5]{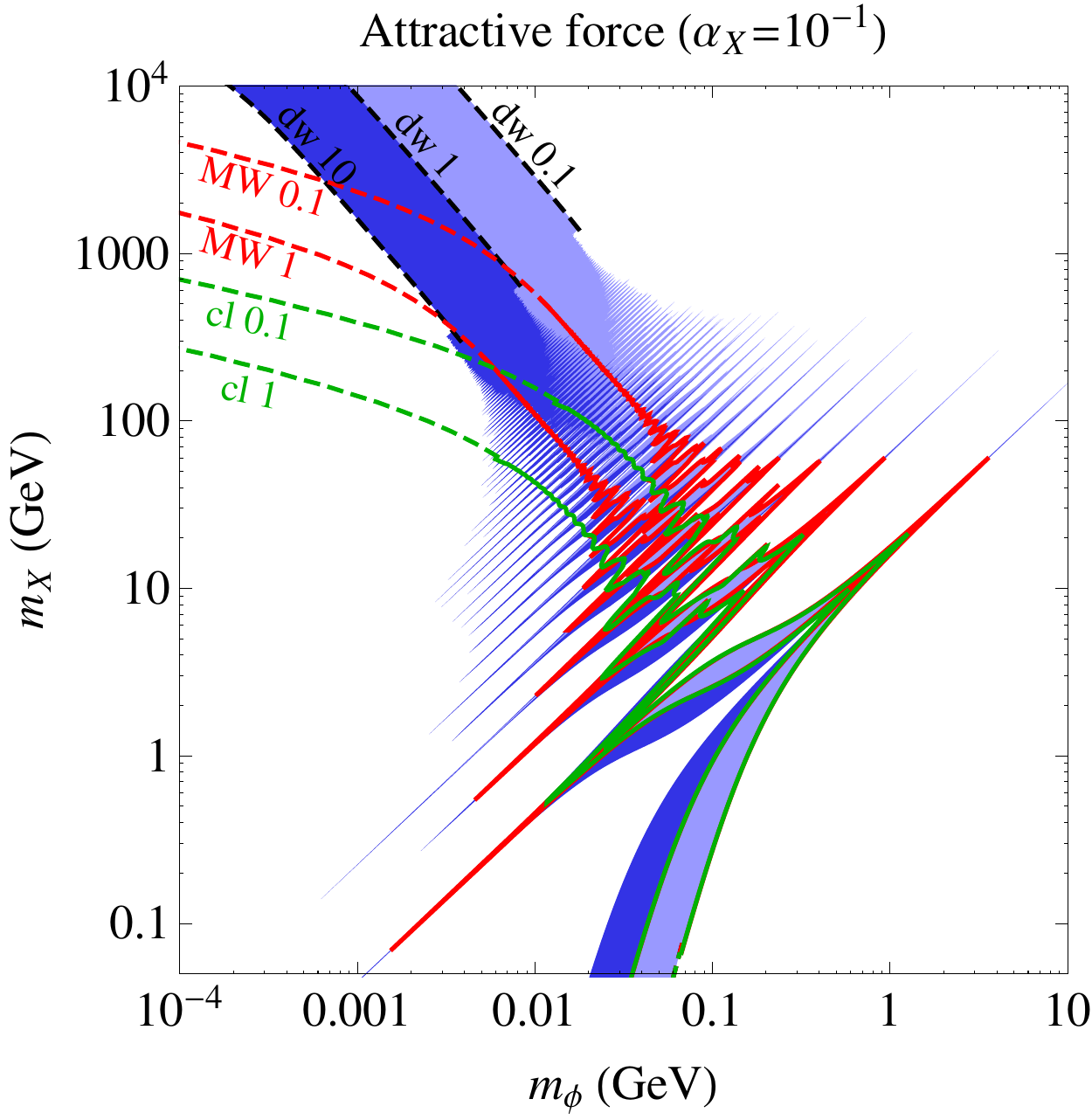} \includegraphics[scale=0.5]{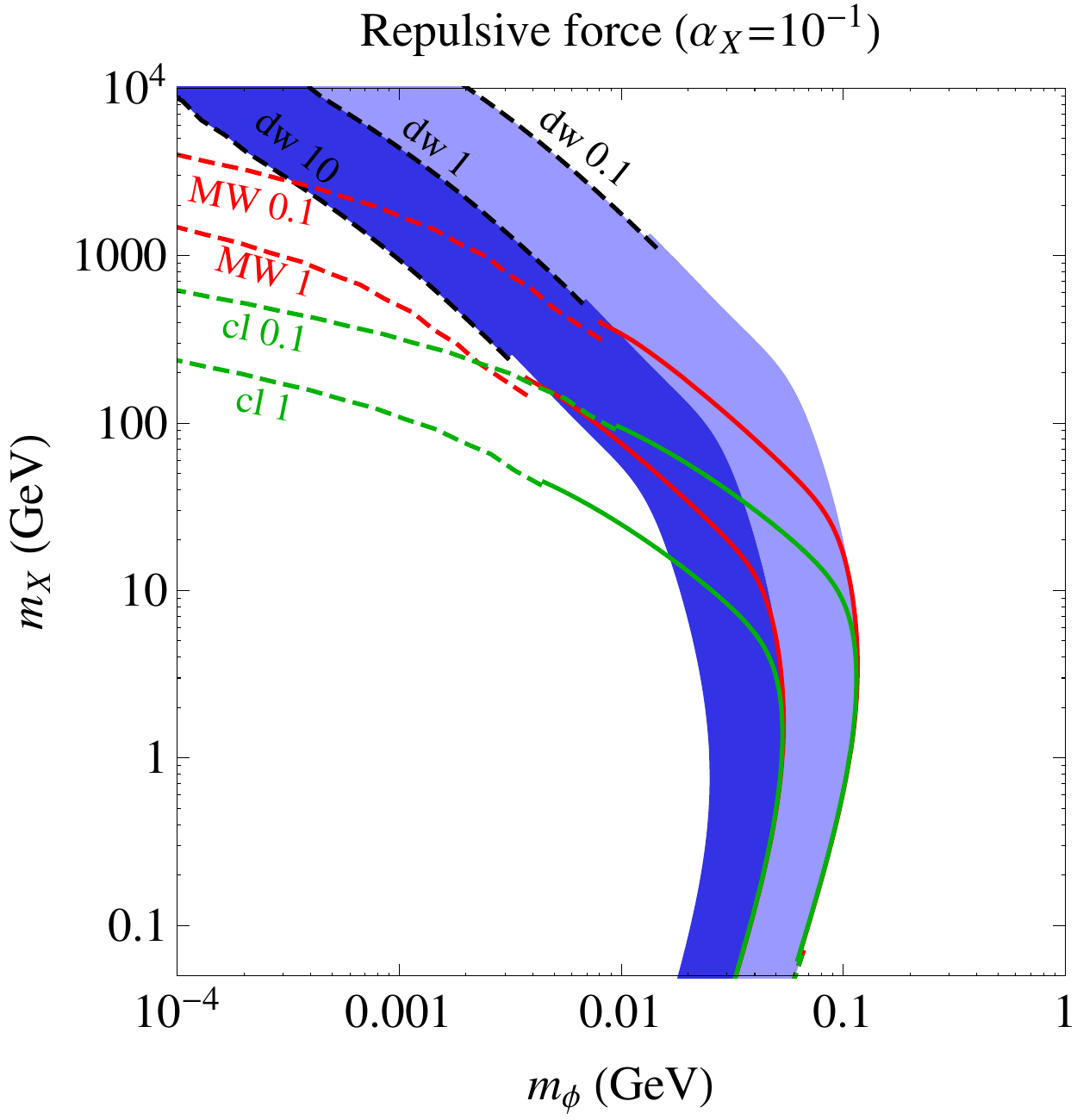}
\includegraphics[scale=0.5]{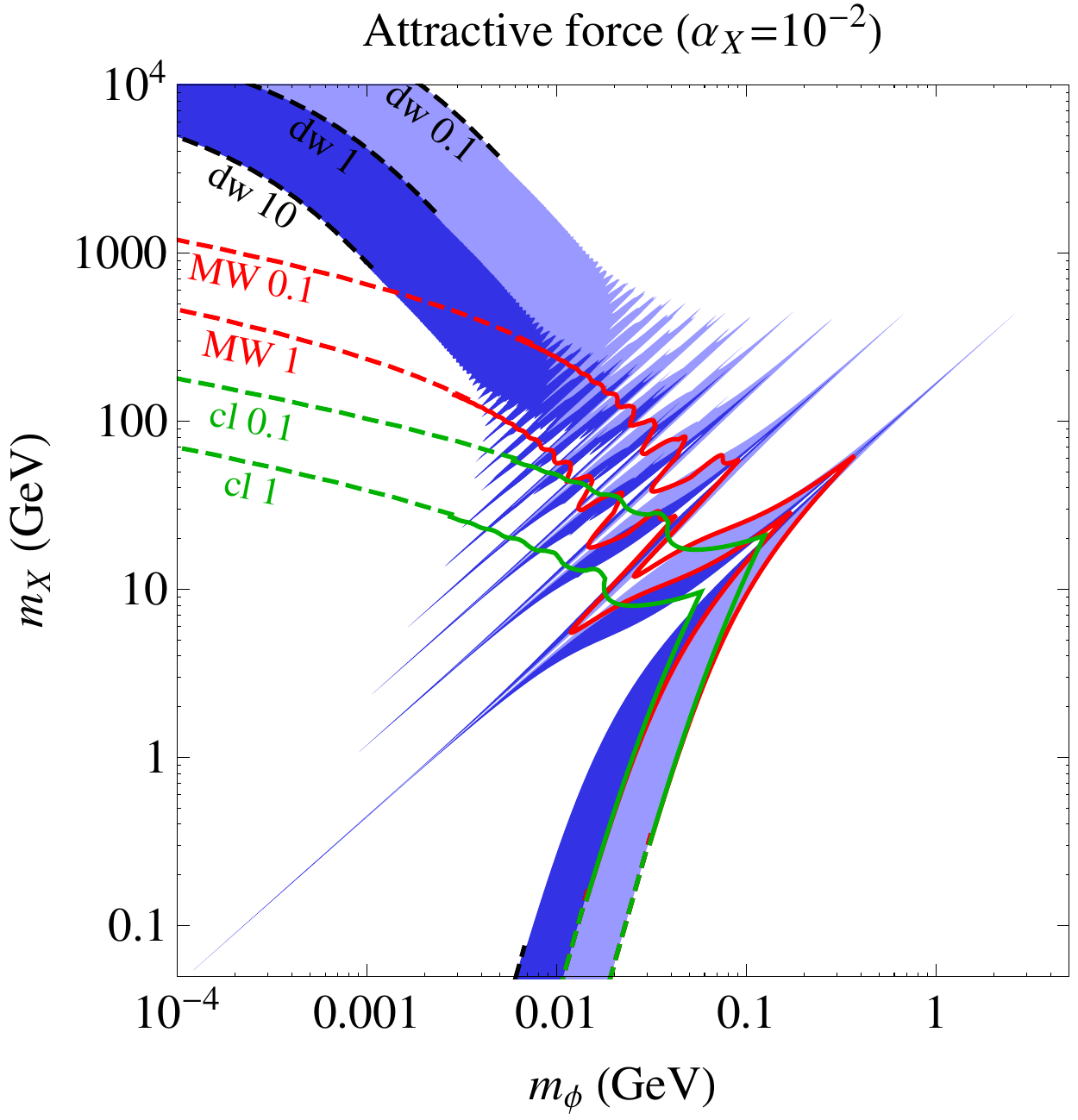} \includegraphics[scale=0.5]{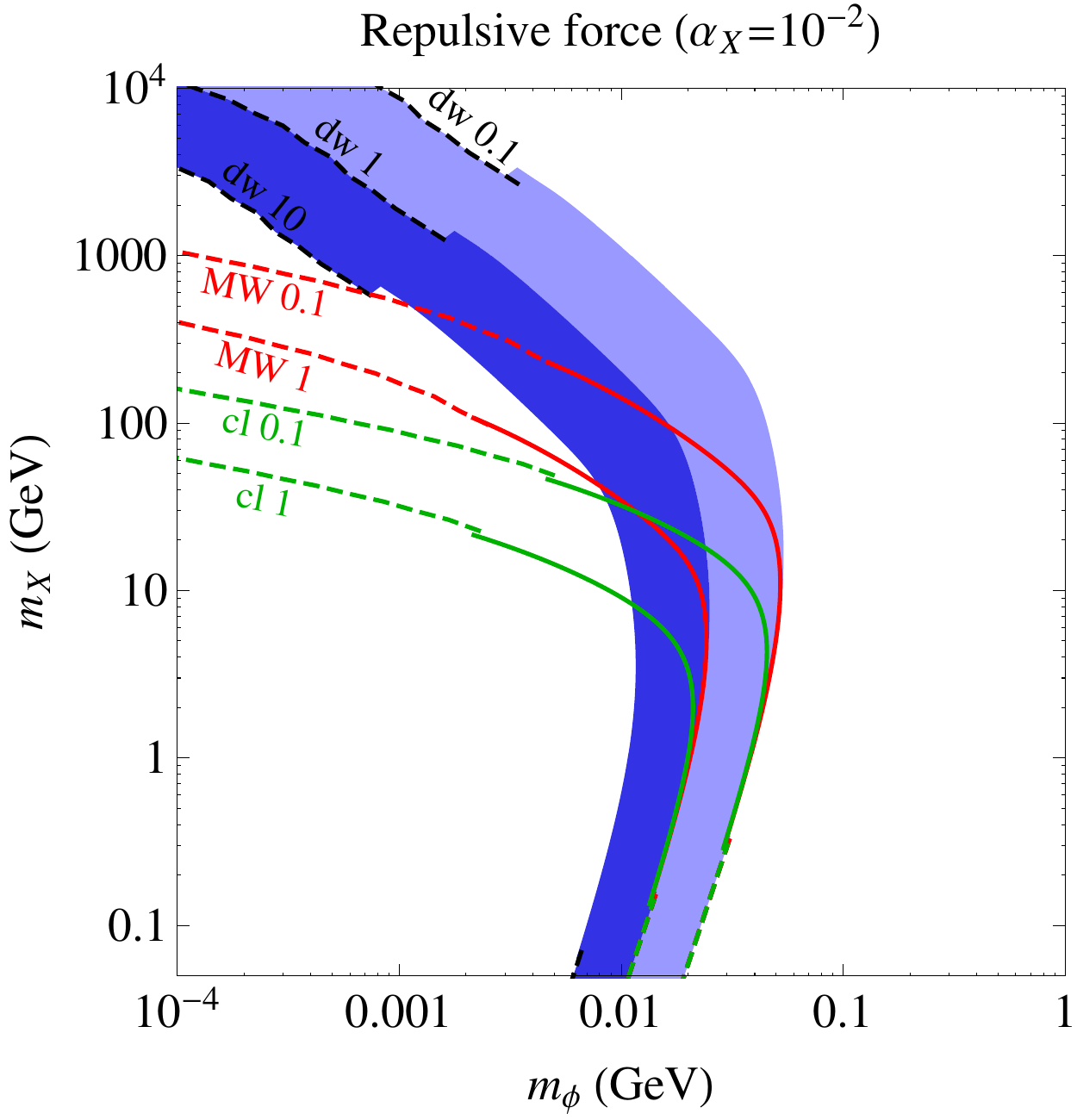}
\includegraphics[scale=0.5]{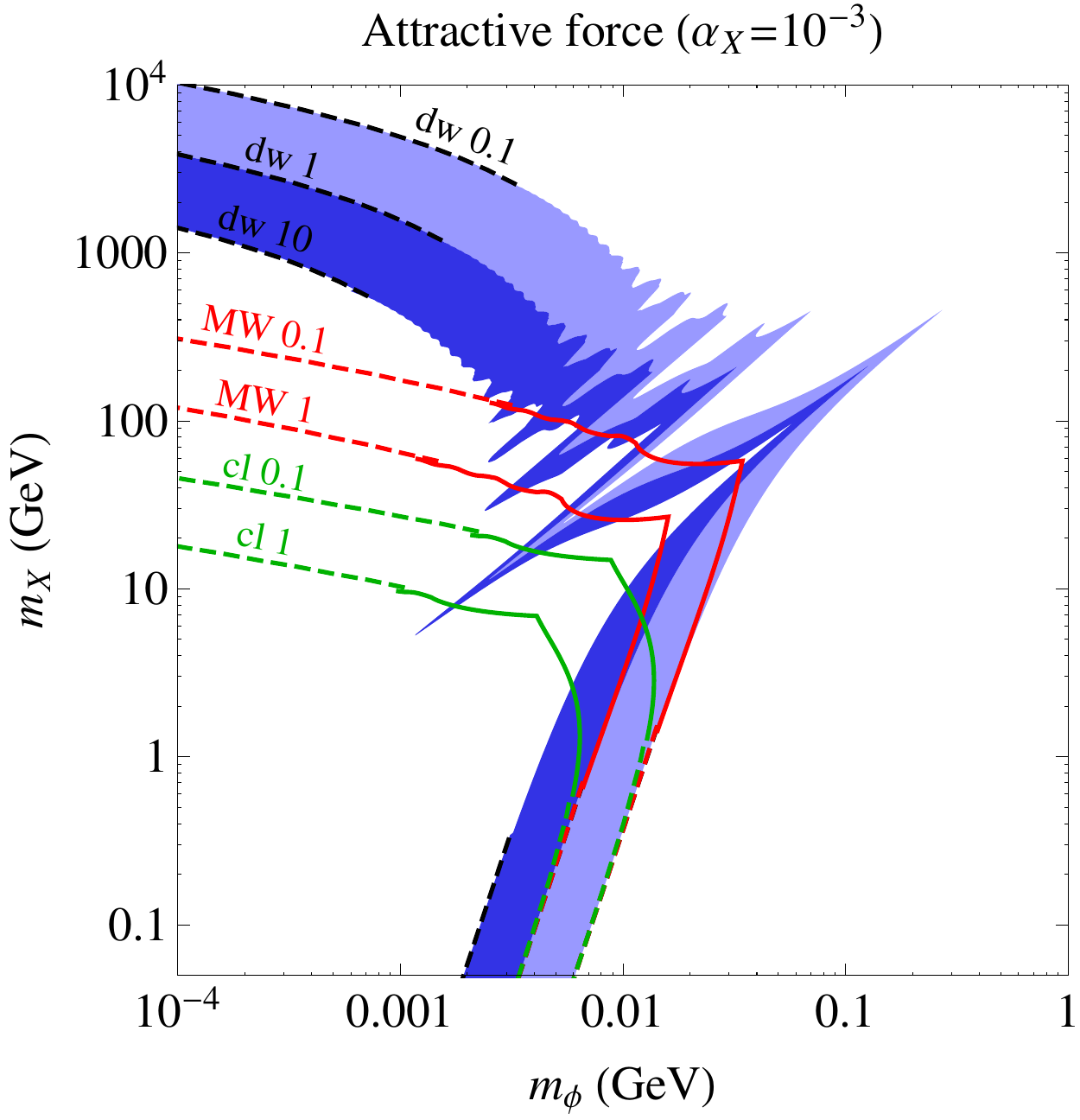} \includegraphics[scale=0.5]{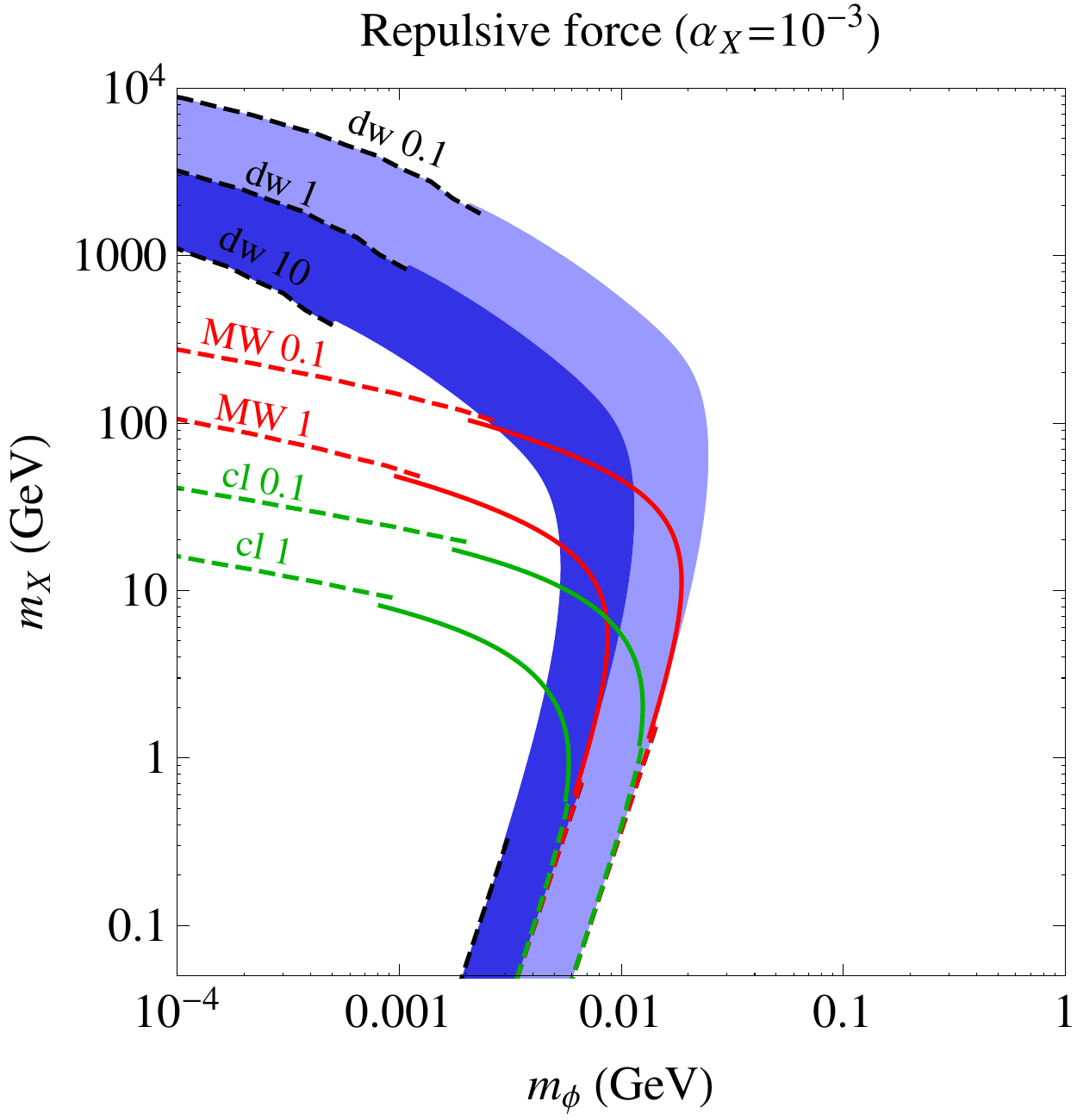}
\caption{Parameter space consistent with astrophysical bounds for attractive (left) and repulsive (right) potentials for different $\alpha_X$.  Blue regions show where DM self-scattering solves small scale structure anomalies, while red (green) show bounds on Milky Way (cluster) scales.  Numerical values give $\langle \sigma_T\rangle /m_X$ in $\cmg$ on dwarf (``dw''), Milky Way (``MW''), and cluster (``cl'') scales.  See text for details.}
\label{ResonantPlots}
\end{figure}

Our results for $\langle \sigma_T\rangle$ are presented in Fig.~\ref{ResonantPlots}.  For both attractive ({left}) and repulsive ({right}) potentials, we show the allowed range of $(m_X,m_\phi)$ for $\alpha_X = 10^{-1}, \, 10^{-2}, \, 10^{-3}$.  Astrophysical bounds on different scales are indicated as follows:
\begin{itemize}
\item Blue regions show $0.1 < \langle \sigma_T/m_X \rangle < 1 \; \cmg$ (light) and $1 < \langle \sigma_T \rangle < 10 \; \cmg$ (dark) on dwarf scales ($v_0 = 10 \; \kms$), required for solving small scale structure anomalies.  
\item Red contours show $\langle \sigma_T \rangle/m_X = 0.1$ and $1 \; \cmg$ on MW scales ($v_0 = 200 \; \kms$).
\item Green contours show $\langle \sigma_T \rangle/m_X = 0.1$ and $1 \; \cmg$ on cluster scales ($v_0 = 1000 \; \kms$).
\end{itemize}
The dashed lines indicate where we use analytic formulae for $\sigma_T$, given in Eq.~(\ref{plasma}), to interpolate our results into the classical (top) and Born (bottom) regimes.  The fact that our numerically computed contours match well onto these regimes demonstrates the consistency between the numerical and analytic results.

Since N-body simulations have been performed for only a limited choice of cross sections, the precise numerical values of these constraints are open to interpretation.  For a constant cross section, Ref.~\cite{Rocha:2012jg} found that $\sigma_T/m_X = 1 \; \cmg$ is too large, causing too-small central densities in dwarf spheroidals and clusters and is marginally consistent with ellipticity constraints on MW scales, while $\sigma_T/m_X = 0.1 \; \cmg$ satisfies all constraints including on dwarf scales.  On the other hand, simulations with a velocity-dependent cross section (assuming a classical, attractive form for $\sigma_T$) have favored larger values on dwarf scales,  $\sigma_T/m_X \sim 10 \; \cmg$~\cite{Vogelsberger:2012ku}.  Therefore, we expect that the actual astrophysical bound on MW (cluster) scales lies between the red (green) lines between $\langle \sigma_T \rangle/m_X = 0.1 - 1 \; \cmg$, with the area to the left excluded, while the blue regions are favored by solving small scale structure anomalies.  More precise limits require future N-body simulations utilizing the full velocity-dependent form for $\sigma_T(v)$, as a function of the DM parameters.

The most striking features emerging from our numerical calculation are the pattern of quantum mechanical resonances and antiresonances for the attractive potential case (absent for the repulsive case).  For fixed $\langle\sigma_T\rangle/m_X$, the (anti)resonances favor larger (smaller) $m_X$, corresponding to peaks pointing to the upper right (lower left) in Fig.~\ref{ResonantPlots}.   These features are more pronounced for smaller $v$ and larger $\alpha_X$ since the conditions $m_X v/m_\phi \lesssim 1$  and $\alpha_X m_X/m_\phi \gtrsim 1$ govern the onset of quantum mechanical and non-perturbative effects, respectively.  It is clear that the resonant regime corresponds to a large region of parameter space, $m_X \sim {\rm GeV} - {\rm TeV}$, where $\sigma_T$ is computed numerically. In the next section, we will derive an analytical formula for $\sigma_T$ in the resonant regime.
%, where $\sigma_T$ must be computed numerically.  Nevertheless, the particular locations these resonances can be understood analytically (see below).

Our main conclusion from Fig.~\ref{ResonantPlots} is that for a wide range of $(\alpha_X, m_X)$, self-interacting DM can explain small scale structure anomalies while remaining consistent with other astrophysical bounds.  
\begin{itemize}
\item A wide range for the DM mass $m_X$ is allowed, from sub-GeV to multi-TeV or beyond.
\item A wide range of perturbative couplings $\alpha_X$ are allowed; we explicitly showed results for $\alpha_X$ between $10^{-1}$ and $10^{-3}$.
\item For fixed $(m_X,\alpha_X)$, the mediator mass is determined within an order of magnitude by astrophysical bounds.  Generally, for $m_X < {\rm TeV}$, we require $m_\phi \sim 1- 100$ MeV, with smaller $m_\phi$ for $m_X > {\rm TeV}$.
\end{itemize}
Future observations on MW and cluster scales can play a key role in narrowing this parameter space by giving additional velocity data points for $\sigma_T(v)$.  For example, evidence for self-interactions on larger scales at the level of $\langle \sigma_T\rangle/m_X \sim 0.1 \; \cmg$ would favor light DM at the GeV-scale; excluding self-interactions below this level would favor heavier DM.

%%%%%%%%%%%%%%%%%%%%%%%%%%%%%%%%%%%%%%%%%%%
\section{Resonant $s$-wave scattering: analytic results}
%%%%%%%%%%%%%%%%%%%%%%%%%%%%%%%%%%%%%%%%%%%5
\label{sec:para}
We derive a new analytic formula for the $s$-wave scattering cross section that is valid in the resonant regime.  This result provides an accurate description of DM scattering in a parameter region which has not been previously analytically accessible and is complementary to the Born and classical regimes.  We give a simple analytic condition for resonances and antiresonces to occur, and we confirm our results against our numerical computation. 

Although the Schr\"{o}dinger equation cannot be solved analytically for the Yukawa potential in the non-perturbative regime, a useful proxy is provided by the Hulth\'{e}n potential
\be \label{hulthpot}
V(r) = \pm  \frac{\alpha_X \delta \, e^{- \delta r}}{1 - e^{-\delta r}} \; ,
\ee
which is analytically solvable for $\ell = 0$.  The Yukawa and Hulth\'{e}n potentials behave similarly, scaling as $1/r$ at short distances and becoming screened for large distances.  The Hulth\'{e}n screening mass $\delta$ is assumed to be related to $m_\phi$ by $\delta = \kappa m_\phi$, where $\kappa$ is an $\mathcal{O}(1)$ numerical constant.  In computing the Sommerfeld enhancement for DM annihilation, Ref.~\cite{Cassel:2009wt} showed that Eq.~\eqref{hulthpot} provides an accurate analytic approximation of the Yukawa potential, with $\kappa = \pi^2/6$.  Here, we follow a similar analysis to compute the cross section for DM scattering; however, we keep $\kappa$ as a free parameter.

Defining $c\equiv \alpha_X m_X/\delta$ and substituting the Hulth\'{e}n for Yukawa potentials, Eq.~\eqref{radial2} becomes 
\be \label{radial3}
\left( \frac{ d^2}{dx^2} + a^2 \mp \, \frac{c^{-1} e^{- x/c }}{1-e^{-x/c}} \right) \chi_0(x) = 0 \; ,
\ee
for $\ell=0$.  With another change of variables $t = 1 - e^{-x/c}$ and $\chi_0(x) = t(1-t)^{i a c} f(t)$, Eq.~\eqref{radial3} can be expressed as Euler's hypergeometric differential equation
\be \label{hyper}
\left( t(1-t) \frac{d^2 }{dt^2} + \big[ 2 -  (\lambda_+ + \lambda_- + 1) t\big] \,\frac{ d}{dt} - \lambda_+ \lambda_- \right) f(t) = 0 \, ,
\ee
with solution \mbox{$f(t) = \!\,_2 F_1(\lambda_+, \lambda_-;  2; t)$}, and where the coefficients $\lambda_\pm$ are defined by
\be
\lambda_\pm = \left\{ \begin{array}{ll} 1 + i a c \pm i \sqrt{ c + a^2 c^2  } & \;\; {\rm repulsive \; potential} \\
1 + i a c \pm \sqrt{ c - a^2 c^{2}} &\;\; {\rm attractive \; potential} \end{array} \right. \, .
\ee
Thus, the full solution is $\chi_0 = t (1-t)^{i a c} \,_2 F_1(\lambda_+, \lambda_-;  2; t)$, up to an irrelevant normalization.

To compute the phase shift $\delta_0$, we are interested in the behavior of $\chi_0$ as $x \to \infty$ (or $t\to 1$).  In this limit, we have\footnote{This follows using the identity
\begin{align*}
 \!\,_2 F_1(A, B; C; t) \;= \;&\frac{\Gamma(C) \Gamma(C-A-B)}{\Gamma(C-A) \Gamma(C-B)}  \,_2 F_1(A,B; A+B-C+1; 1-t) \\
&  +  \frac{\Gamma(C) \Gamma(A+B-C)}{\Gamma(A) \Gamma(B)}\, (1-t)^{C-A-B} \,_2 F_1(C-A,C-B; C-A-B+1;1- t) \, ,
\end{align*}
which is valid for non-integer $A+B-C$, and also using $\!\,_2 F_1(A,B; C;0) = 1$.
 }
\be \label{chiinf}
\chi_0(x) \underset{x \to \infty}{\longrightarrow} \frac{\Gamma(\lambda_+ + \lambda_- - 2)}{\Gamma(\lambda_+) \Gamma(\lambda_-)} \, e^{i a x} + \frac{\Gamma( 2 -\lambda_+ - \lambda_- )}{\Gamma(2- \lambda_+) \Gamma(2- \lambda_-)} \, e^{-i a x} \, \propto  \sin(a x + \delta_0) \, ,
\ee
where the phase shift is given by
\be \label{deltaell}
\delta_0 = \arg \left( \frac{ i \Gamma( \lambda_+ + \lambda_- - 2)}{\Gamma(\lambda_+) \Gamma(\lambda_-)} \right) \; .
\ee
To the extent that $s$-wave scattering dominates, we expect $\sigma_T \approx 4\pi \sin^2 \delta_0/k^2$ to be a useful analytic approximation to the full numerical calculation.  On the other hand, when $m_X v/m_\phi \gtrsim 1$, $\ell>0$ partial waves become important and our analytic result is no longer valid.\footnote{Ref.~\cite{Cassel:2009wt} generalized this method to $\ell >0$ by approximating the centrifugal term by a different function allowing a solution to the Schr\"{o}dinger equation.  However, the modified centrifugal term alters the long distance behavior of the wave function, and the $\ell > 0$ phase shifts we obtain by this method do not agree with our numerical calculations.}  

The existence of $s$-wave resonances can be inferred from Eq.~\eqref{deltaell} by considering the zero velocity limit (since $s$-wave resonances correspond to bound states at zero energy).  First, we consider the attractive case.  Expanding Eq.~\eqref{deltaell} for small $a$ (recall $2a = v/\alpha_X$), we have
\be \label{expand}
\delta_0 \underset{v \to 0} \longrightarrow - \big[ 2 \gamma + \psi(1+ \sqrt{c}) + \psi(1- \sqrt{c}) \big]  a c
\ee
with digamma function $\psi(z) = \Gamma^\prime(z)/\Gamma(z)$ and Euler-Mascheroni constant $\gamma$.  Thus, as $v \to 0$, the phase shift scales as $\delta_0 \propto v$ and $\sigma_T$ approaches to a constant.  However, this expansion breaks down when $\sqrt{c}=n$, where $n$ is a positive integer, due to poles in the gamma function.  In this case, Eq.~\eqref{deltaell} gives a maximal phase shift $\delta_0 \to \pm \pi/2$ for $v \to 0$, corresponding to a resonance where the cross section is enhanced as $\sigma_T \propto 1/v^2$.  In terms of physical parameters, the resonance condition is
\be \label{rescond}
\frac{\alpha_X m_X}{\kappa  m_\phi} = n^2 \, , \quad n = 1, \, 2, \, 3, \, ...
\ee
As expected, this is the same resonance condition derived for Sommerfeld enhancements~\cite{Cassel:2009wt}, since the same bound state formation is relevant for both scattering and annihilation.  We also note the appearance of antiresonances ($\delta_0 = 0$), with vanishing $s$-wave cross section.  From Eq.~\eqref{expand}, the antiresonance condition is
\be
\frac{\alpha_X m_X}{\kappa m_\phi} = r^2 \, , \quad r \approx 1.69, \, 2.75, \, 3.78, \, 4.80, \, 5.81, \, ...
\ee
where $r$ corresponds to positive roots of the equation $2\gamma+\psi(1+r) + \psi(1-r) = 0$.  On the other hand, for a repulsive potential, we have
\be  \label{expand2}
\delta_0 \underset{v \to 0} \longrightarrow - \big[ 2 \gamma + \psi(1+ i \sqrt{c}) + \psi(1- i \sqrt{c}) \big]  a c \; .
\ee
As expected, there is no possibility of resonances, since poles of the gamma function are along the real axis only, nor antiresonances, since the quantity in brackets is strictly positive.

The numerical value of $\kappa$ can be determined {\it a posteriori}.  In computing the Sommerfeld-enhanced annihilation cross section, Ref.~\cite{Cassel:2009wt} fixed $\kappa =\pi^2/6 \approx 1.64$ in order to match the perturbative result in the Born limit at zero velocity.  Applying this prescription to scattering, we wish to relate the Born cross section in Eq.~\eqref{born} to our result from the Hulth\'{e}n potential for $v \to 0$.  In the perturbative limit, Eqs.~\eqref{expand} and \eqref{expand2} give $\delta_0 = \pm 2\zeta(3)  ac^2$, and we have
\be
\sigma_T^{\rm Born} = \frac{4\pi  \alpha_X^2 m_X^2}{m_\phi^4} \, , \quad 
\sigma_T^{\rm Hulth\acute{e}n} = \frac{16 \pi  \alpha_X^2 m_X^2 \zeta(3)^2 }{\kappa^4 m_\phi^4} \; .
\ee
Equating these cross sections gives $\kappa =  \sqrt{2 \zeta(3)} \approx 1.55$.\footnote{The small difference in $\kappa$ stems from a difference in matching the Yukawa and Hulth\'{e}n wavefunctions at $r \to 0$ or $r \to \infty$.  Ref.~\cite{Cassel:2009wt} obtains $\kappa =\pi^2/6$ by equating the wavefunctions at $r \to 0$, requiring that the integral $\int^\infty_0 dr^\prime \, r^\prime\, V(r^\prime)$ is matched between the Yukawa and Hulth\'{e}n potentials, using the Lippmann-Schwinger equation.  Following the same argument, but for $r \to \infty$, one requires $\int^\infty_0 dr^\prime \, {r^\prime}^2\, V(r^\prime)$, giving our result $\kappa = \sqrt{2 \zeta(3)}$.}  However, since there is no unique exact value for $\kappa$ outside the Born limit, we take simply $\kappa = 1.6$ which provides an accurate choice across a wide parametric range.

\begin{figure}
\includegraphics[scale=0.64]{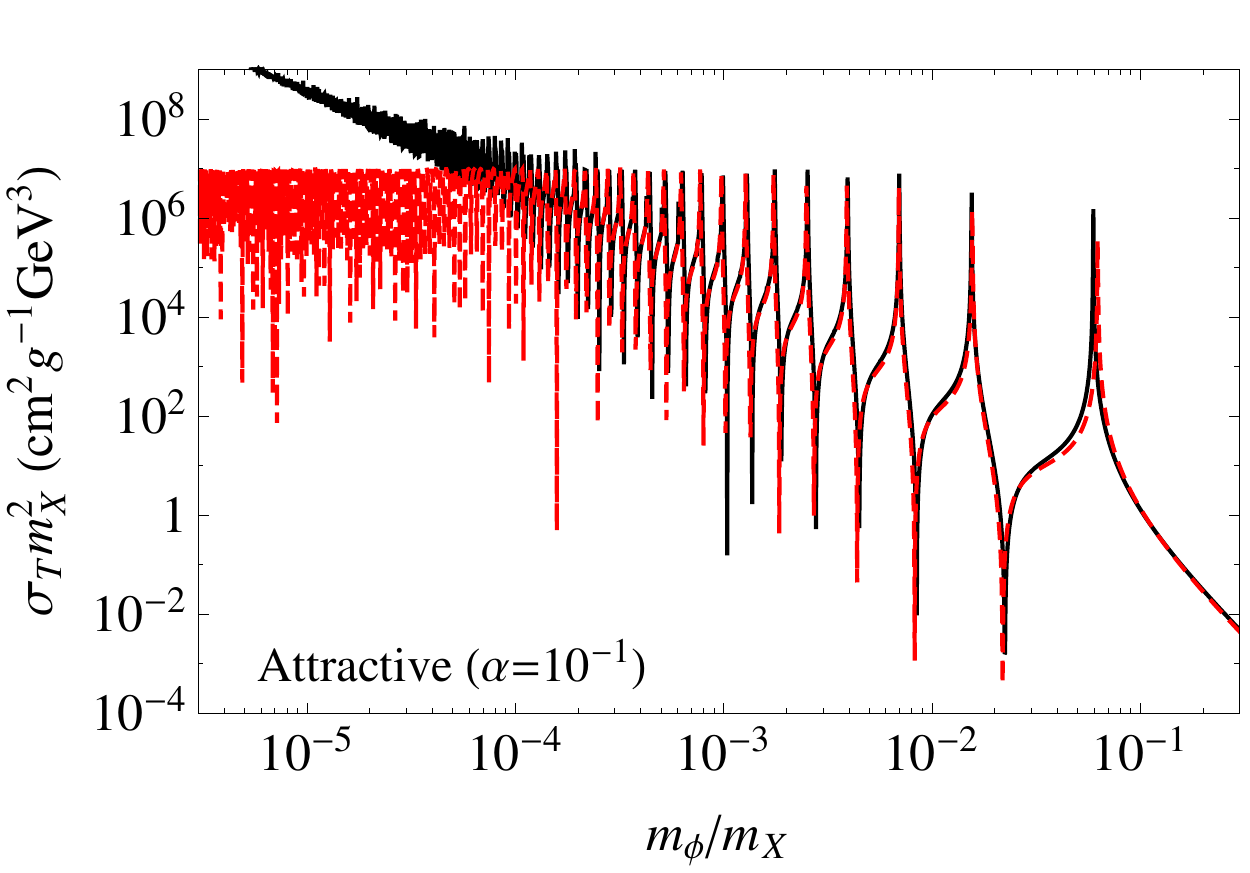} \includegraphics[scale=0.64]{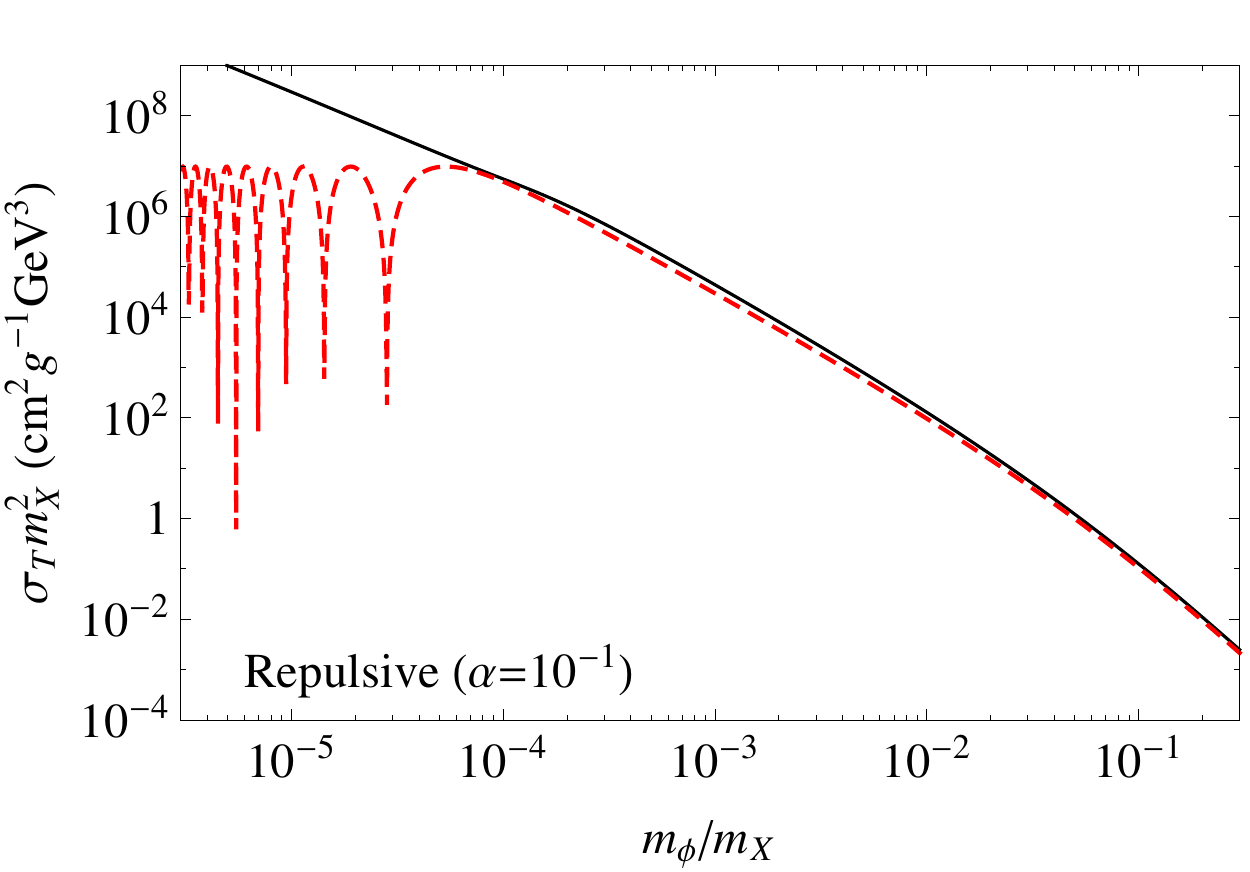}
\includegraphics[scale=0.64]{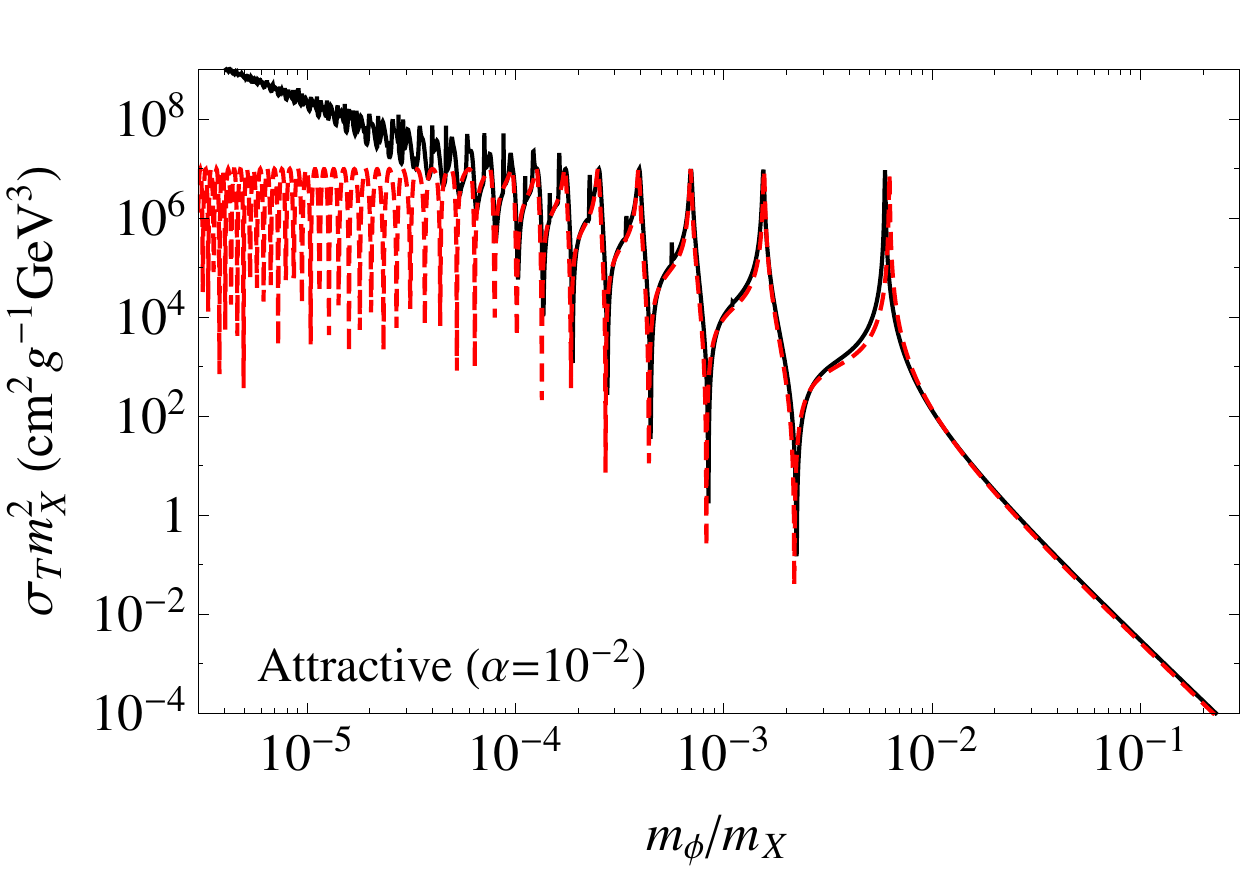} \includegraphics[scale=0.64]{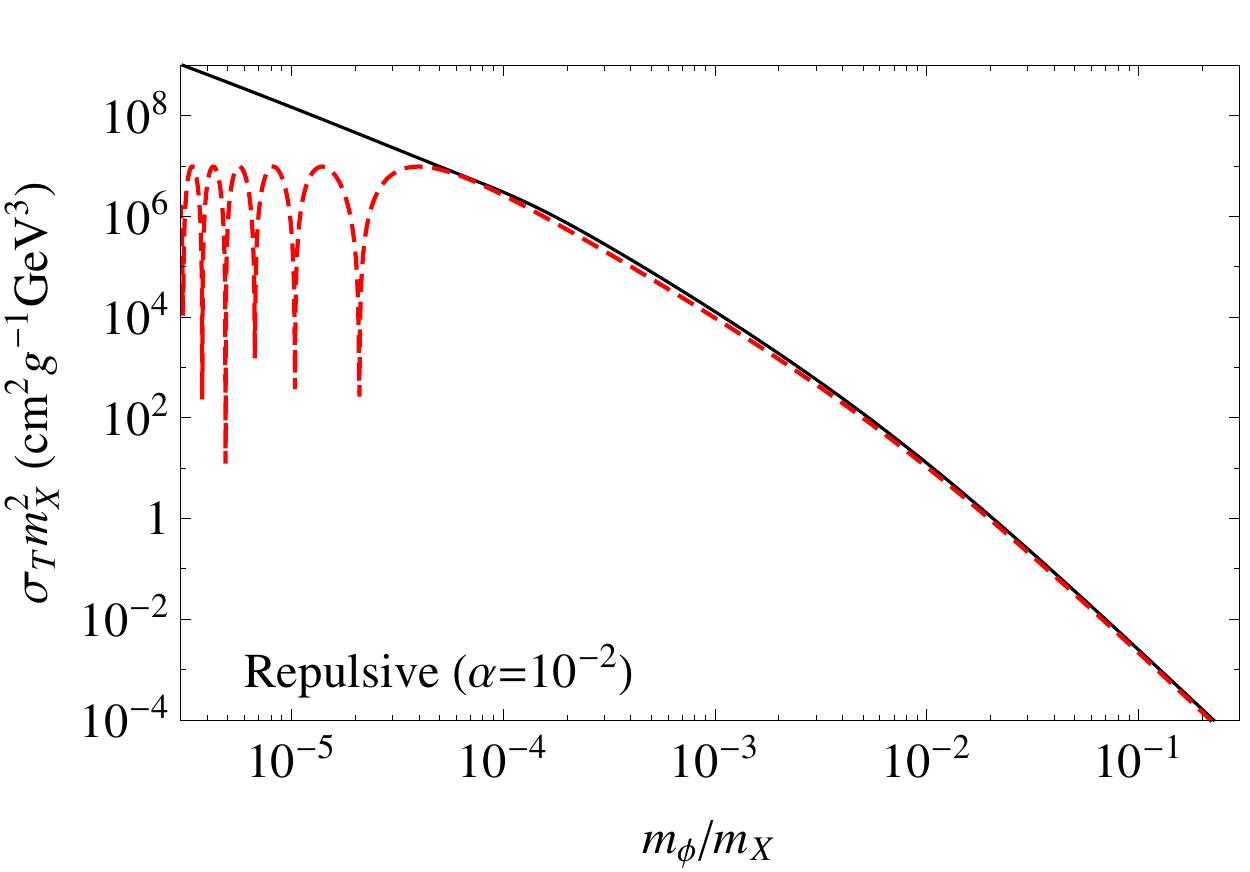}
\includegraphics[scale=0.64]{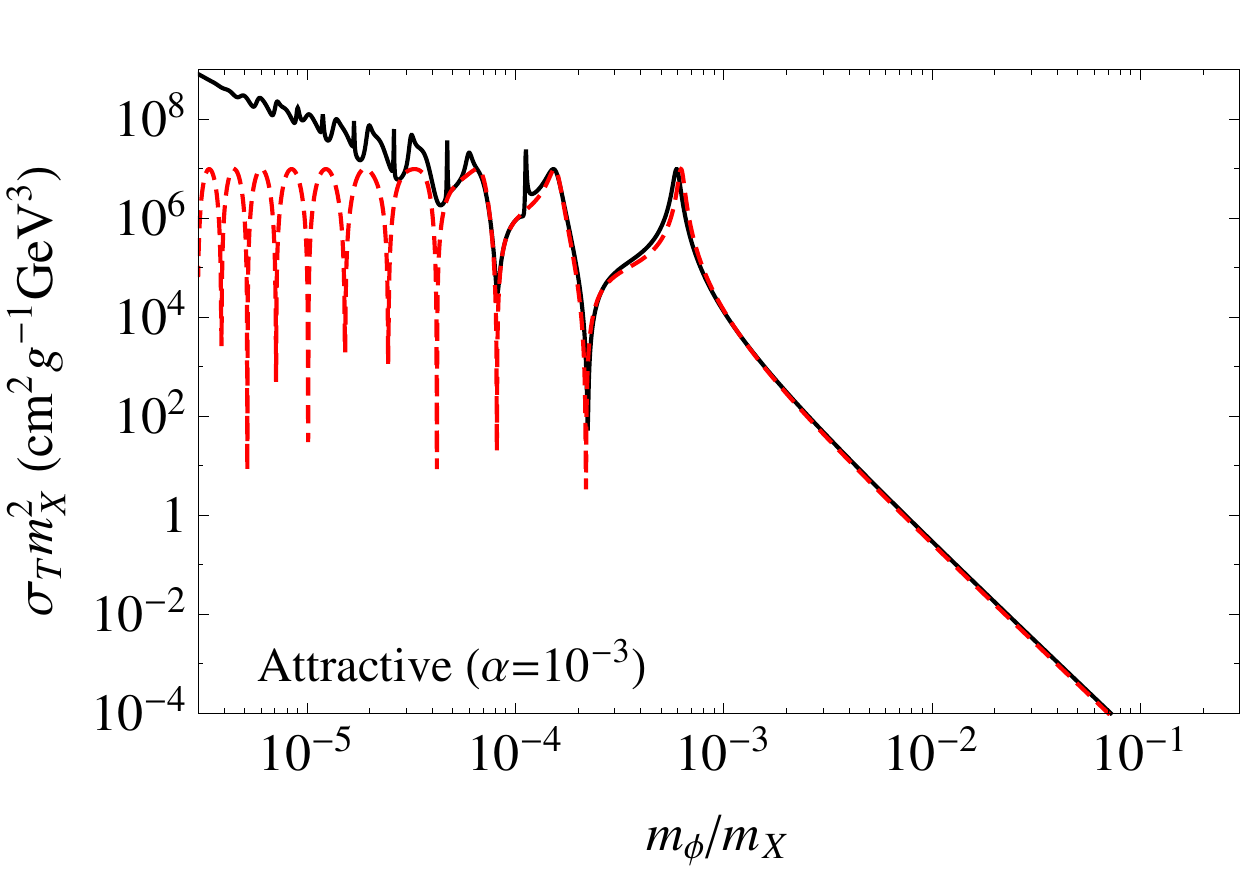} \includegraphics[scale=0.64]{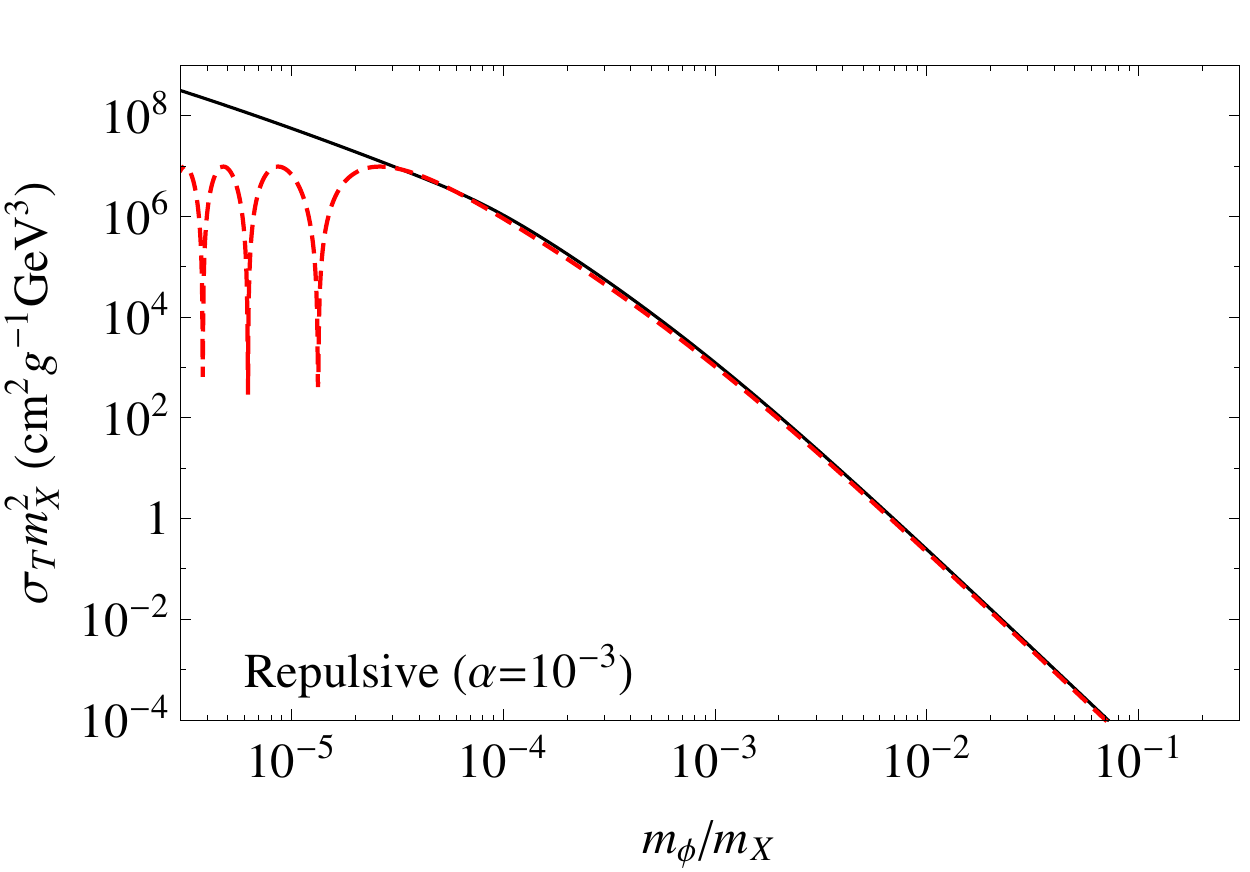}
\caption{Numerical calculation (black solid) and analytic $s$-wave result (red dashed) for $\sigma_T m_X^2$ as a function of $m_\phi/m_X$, for $v=10$ km/s, $\alpha_X = 10^{-3}-10^{-1}$, and both attractive and repulsive interactions.  The analytic approximation breaks down for $m_X v /m_\phi \gtrsim 1$, when $\ell >0$ partial waves become important.}
\label{AnalyticPlots}
\end{figure}

Next, we compare our analytic result for $\sigma_T$ with our numerical calculation, shown in Fig.~\ref{AnalyticPlots}.  Taking a typical dwarf velocity $v=10$ km/s, we plot $\sigma_T m_X^2$ as a function of $m_\phi/m_X$, calculated numerically (black solid) and analytically from the Hulth\'{e}n potential, with $\kappa=1.6$ (red dashed).  Each panel shows a different coupling, $\alpha_X = 10^{-1}, \, 10^{-2}, \, 10^{-3}$, for either an attractive (left) or repulsive (right) interaction.\footnote{The quantity $\sigma_T m_X^2$ is useful to consider since it depends only on $m_\phi/m_X$, after $\alpha_X$ and $v$ are fixed, rather than $m_X$ and $m_\phi$ separately.  Thus, for every point along the curves shown in Fig.~\ref{AnalyticPlots}, one can fix $\sigma_T/m_X$ to any desired value ({\it e.g.}, $1 \cmg$) by taking the appropriate values of $m_X,m_\phi$.}   Our numerical and analytic results agree remarkably well for $m_\phi/m_X \gtrsim v \approx 3\times 10^{-5}$, where scattering is predominantly $s$-wave, accurately mirroring the pattern of resonances and antiresonances within the resonant regime (for the attractive case).  This agreement provides a highly nontrivial confirmation of our numerical calculation.  For $m_\phi/m_X \lesssim v$, the results diverge as $\ell >0$ partial waves become more important, as expected.

Lastly, we provide a series of benchmark parameters for resonant $s$-wave scattering.  This case provides a simple and novel velocity-dependence for DM scattering, $\sigma_T \propto v^{-2}$, and it would be interesting to incorporate this case within N-body simulations.  On-resonance ($\delta_0 = \pi/2$), the differential cross section is $d\sigma/d\Omega = (m_X v/2)^{-2}$, giving $\sigma_T = 16\pi m_X^{-2} v^{-2}$.  In Table~I we list benchmark parameters that give resonant scattering and also produce the correct DM relic density (via $p$-wave annihilation $X \bar X \to \phi \phi$, see Sec.~\ref{sec:relic}).  We consider several values of $m_X$ to fix $\sigma_T/m_X$ on dwarf scales; the remaining parameters $(m_\phi,\alpha_X)$ are determined by resonance condition, Eq.~\eqref{rescond} with $n=1$ and $\kappa \approx 1.7$, and the relic density.  For these parameter points, we have checked that $m_X v/m_\phi \lesssim 1$ up to cluster scales $v \sim 1000 \, \kms$, validating neglect of $\ell > 0$ partial waves.

%%%%%%%%%%%%%%%%%%%%%%%%%%%%%%%%%%%%%%%%%%%5
\begin{table}[t!]
\begin{center}
\begin{tabular}{cccc}
\hline
$\sigma_T/m_X$ at $v=10 \, \kms$ & $m_X$ & $m_\phi$ & $\alpha_X$ \\
\hline
$1 \, \cmg$ & $210$ GeV & $2.8$ GeV & $2.3 \times 10^{-2}$ \\
$10 \, \cmg$ & $100$ GeV & $0.67$ GeV & $1.1 \times 10^{-2}$ \\
\hline
\end{tabular} \label{benchtab}
\caption{Benchmark points for resonant $s$-wave scattering with $d\sigma/d\Omega = (m_X v/2)^{-2}$ and $\sigma_T = 16\pi (m_X v)^{-2}$, consistent with correct DM relic density.}
\end{center}
\end{table}
%%%%%%%%%%%%%%%%%%%%%%%%%%%%%%%%%%%%%%%%%%%

\section{Relic density}
\label{sec:relic}

In the above discussion, we have taken $\ax$ to be a free parameter. In this section, we fix $\ax$ for a given $(m_\phi,~m_X)$ through the DM relic density, set by the annihilation process $X\bar{X}\rightarrow\phi\phi$.  We consider here two representative cases where $\phi$ is a vector or scalar field. The annihilation cross sections at tree-level are 
\be
\left(\sigma_{\rm an} v\right)^{\rm tree}_V=\frac{\pi\ax^2}{\mx^2}\sqrt{1-\frac{\mphi^2}{\mx^2}},~\left(\sigma_{\rm an} v\right)^{\rm tree}_S=\frac{3}{4}\frac{\pi\ax^2}{\mx^2}v^2\sqrt{1-\frac{\mphi^2}{\mx^2}}
\label{eq:ann}
\ee
for the vector and scalar mediators. It is clear that DM annihilation to scalar mediators is a $p$-wave process. Since the mediators have masses around $1-100$ MeV, they will also lead to Sommerfeld enhancements for DM annihilation~\cite{ArkaniHamed:2008qn,Hisano:2004ds}. These enhancements can be important in the early Universe for heavy DM.

The formalism for the symmetric freeze-out with $s$-wave Sommerfeld enhancements has been discussed~\cite{Feng:2010zp}. Here, we expand it to include the $p$-wave case.\footnote{See also~\cite{Chen:2013bi}. } The coupled Boltzmann equations for the species $X$ and $\bar{X}$ can be written as
\be
\frac{dY_{X,\bar X}}{dx}=-\sqrt{\frac{\pi}{45}}m_{\rm pl}m_X\frac{g_{*s}/\sqrt{g_*}}{x^2}\left<\san v\right>(Y_X Y_{\bar X}-Y_{\rm eq}^2),
\ee
where we take the standard definitions\footnote{The reader should not be confused with $x\equiv\ax\mx r$ defined in Sec.~IV.} $x=\mx/T$ and $Y_{X,\bar X}=n_{X,\bar X}/s$, with $n_{X,\bar X}$ the DM number density,  $s$ the entropy density, and $Y_{\rm eq}$ the equilbrium value of $Y_{X,\bar X}$. In addition $m_{\rm pl}\simeq1.2\times10^{19}~{\rm GeV}$ is the Planck mass, $\left<\san v\right>$ the thermally-averaged annihilation cross section, and $g_{*s}$ and $g_*$ are the relativistic degrees of freedom for entropy and energy density, respectively. 

During freeze-out, DM particles have a high velocity and the Sommerfeld enhancement effect is negligible. Thus, the freeze-out temperature can be estimated as usual~\cite{Kolb:1990vq}
\begin{eqnarray}
\nonumber x_f&\simeq&\ln\left[0.038n(n+1)m_{\rm pl}m_X(g/\sqrt{g_*})\sigma_0\right]\\
&&-\left(n+\frac{1}{2}\right)\ln\left(\ln\left[0.038n(n+1)m_{\rm pl}m_X(g/\sqrt{g_*})\sigma_0\right]\right),
\end{eqnarray}
where  $g=2$ is the number of degrees of freedom of $X$ and $\sigma_0$ is given by the relation $\left<\sigma_{\rm an} v\right>=(T_X/\mx)^n\sigma_0$, where $T_X$ is the DM temperature, and $n$ indicates the annihilation type, {\it i.e.}, $n=0$ and $1$ for $s$-wave and $p$-wave annihilation, respectively. 

After freeze-out, $Y_{\rm eq}$ becomes insignificant. Neglecting $Y_{\rm eq}$, we can solve the Boltzmann equations analytically as $Y_{X,\bar X}(x_s)\simeq{3.79}/({m_{\rm pl}\mx J})$ with
\begin{eqnarray}
J=\int^{x_{\rm kd}}_{x_f}\frac{g_{*s}/\sqrt{g_*}}{x^2}\left<\san v\right>dx+\int^{x_{\rm s}}_{x_{\rm kd}}\frac{g_{*s}/\sqrt{g_*}}{x^2}\left<\san v\right>dx,
\end{eqnarray}
where $x_{\rm kd}$ is the value of $x$ at kinetic decoupling and $x_s$ is its value when DM annihilation becomes insignificant and we may stop the integration. Before kinetic decoupling, DM has the same temperature as the thermal bath $T_X=T$. After kinetic decoupling at $T_{\rm kd}$, the DM velocity distribution may be distorted from Maxwell-Boltzmann in scenarios with Sommerfeld-enhanced annihilation, since annihilations preferentially deplete the low velocity population. But as shown in~\cite{Feng:2010zp}, DM self-interactions mediated by $\phi$ can maintain kinetic equilibrium in the parameter region we are interested in, and in this case, we simply take the Maxwell-Boltzmann distribution with $T_X=T^2/T_{\rm kd}$. 

When the DM distribution is thermal with temperature $T_X$, the thermally-averaged cross section in the nonrelativistic limit is 
\be
\left<\san v\right>= \int \frac{d^3 v}{(2\pi v_0^2)^{3/2}} \, e^{-\frac{1}{2} v^2/v_0^2} \, \san v
\ee 
where $v_0=\sqrt{2T_X/\mx}=\sqrt{2/x_X}$. 
We write the annihilation cross section as $\san v=S(\san v)^{\rm tree}$, where $(\san v)^{\rm tree}$ is the cross section calculated at the tree-level and $S$ is the enhancement factor. Thus, the thermally-averaged annihilation cross section is
\be
\left<\san v\right>=\frac{x^{3/2}_X}{2\sqrt{\pi}}\int S(\san v)^{\rm tree} v^2 e^{-x_X v^2/4}dv.
\ee
In the cases we consider, the tree-level annihilation cross sections are given by Eq.~(\ref{eq:ann}) and the Sommerfeld enhancement factors for $s$-wave and $p$-wave annihilations are
\be
S_s=\frac{\pi}{a}\frac{\sinh(2\pi a c)}{\cosh(2\pi ac)-\cos(2\pi\sqrt{c-(ac)^2})}, \quad S_p=\frac{(c-1)^2+4(ac)^2}{1+4(ac)^2} S_s,
\label{eq:se}
\ee
respectively, where we have used $a=v/2\ax$ and $c=6b/\pi^2=6\ax\mx/\pi^2\mphi$~\cite{Cassel:2009wt}.

\begin{figure}[t]
\includegraphics[scale=0.96]{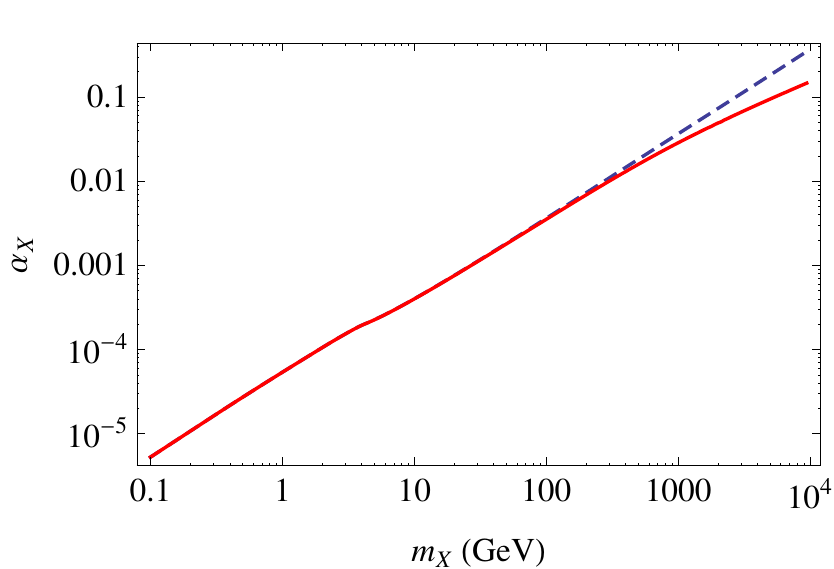}
\includegraphics[scale=0.96]{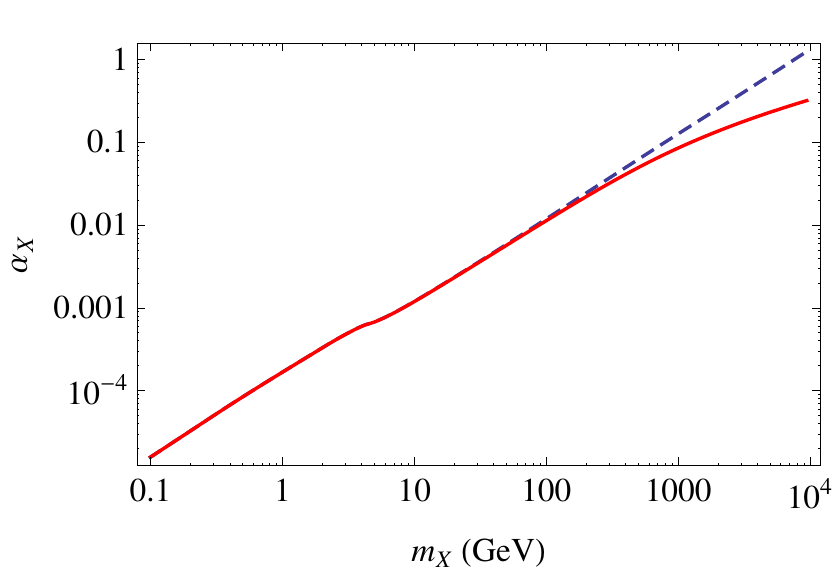}
\caption{The value of $\ax$ required to obtain the correct DM relic density as a function of the DM mass $\mx$ (solid red) for the vector (left) and scalar (right) mediators. We also plot the required $\ax$ (dashed blue) if the Sommerfeld effect is neglected in the early Universe. We take the DM kinetic decoupling temperature $T_{\rm kd}=1~{\rm MeV}$ and the mediator mass $\mphi=10~{\rm MeV}$.   \label{fig:relic}}
\end{figure}

In Fig.~\ref{fig:relic}, we show the value of $\ax$ which gives rise to the observed relic density for the vector (left) and scalar (right) mediators. In the calculation, we have taken the DM kinetic decoupling temperature $T_{\rm kd}=1~{\rm MeV}$ and the mediator mass $\mphi=10~{\rm MeV}$. The Sommerfeld effect in the early Universe can lead to an ${\cal O}(1)$ suppression factor on $\ax$ for $\mx\gtrsim1~{\rm TeV}$, but is negligible for lighter DM. This is because heavier DM requires a larger $\ax$ which results in a larger enhancement factor on DM annihilation in the early Universe. 

Here, we comment on the dependence of the result shown in Fig.~\ref{fig:relic} on $\mphi$ and $T_{\rm kd}$.  Since a large mass hierarchy between $\mx$ and $\mphi$ is required for DM to have sufficient self-interactions to affect structure formation when $\mx\gtrsim1~{\rm TeV}$, the mediator is effectively massless for the Sommerfeld enhancement.  Thus the result is not sensitive to $\mphi$. The value of $\ax$ can also depend on $T_{\rm kd}$. For a small $T_{\rm kd}$, DM particles cool down slowly, which suppresses the Sommerfeld effect. However, typically, this dependence is very mild because the the DM annihilation rate becomes much less than the Hubble expansion rate before the Universe cools to $T_{\rm kd}$, even if the annihilation is enhanced. In our case, we have checked that $\ax$ only changes by less than $3\%$ when we set $T_{\rm kd}$ to be $1~{\rm GeV}$. It is worth noting, however, that $T_{\rm kd}$ may play an important role in the resonance regime. It has been shown that DM can re-couple to the thermal bath after freeze-out in the resonance regime, which leads to a negligible relic density~\cite{Feng:2010zp}. This chemical re-coupling effect only occurs when $T_{\rm kd}$ is high and parameters have to be highly fine-tuned to satisfy the resonance condition exactly. With $T_{\rm kd}=1~{\rm MeV}$, we have checked that chemical re-coupling does not happen and DM has the correct relic density in the resonance regime. 
\begin{figure}[t]
\includegraphics[scale=0.64]{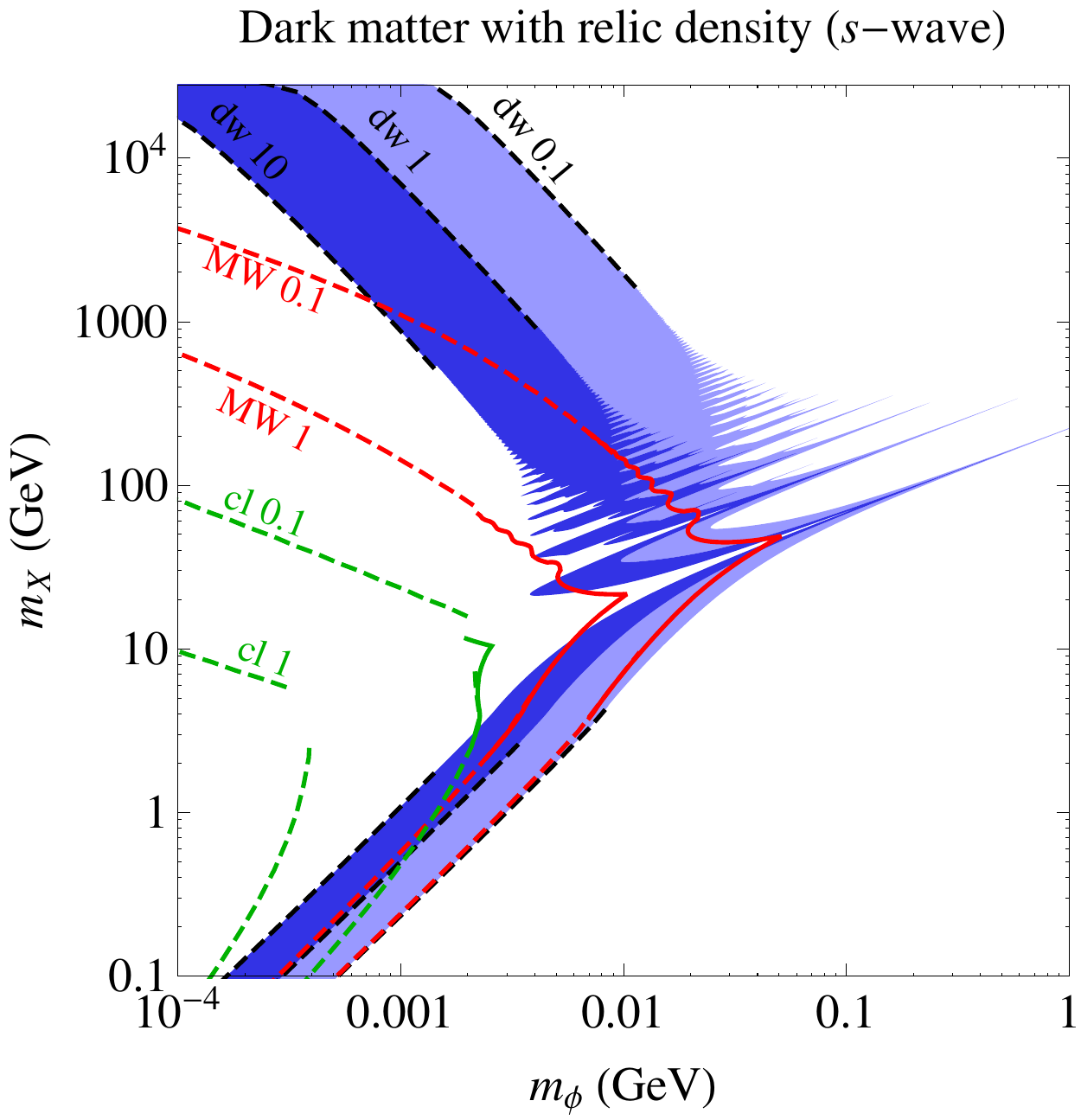}
\includegraphics[scale=0.64]{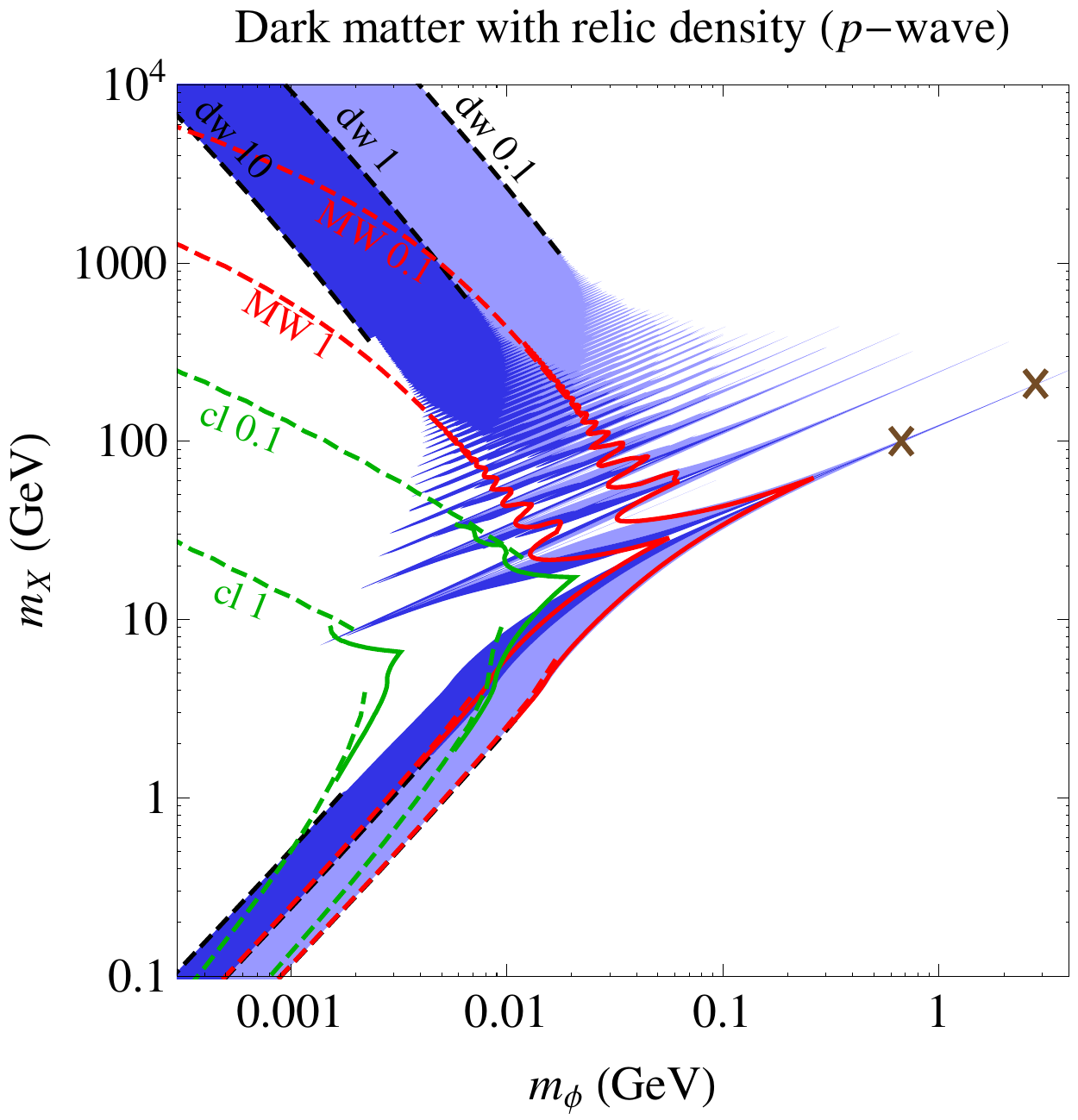}
\caption{Parameter space for self-interacting DM as in Fig.~\ref{ResonantPlots} with $\alpha_X$ fixed to obtain the observed relic density via $X \bar X \to \phi \phi$ annihilation at freeze-out.  The left (right) panel shows the vector (scalar) mediator case whwere annihilation is $s$-wave ($p$-wave).  Crosses show benchmark points in Table~\ref{benchtab}.  The lines and colored regions are as in Fig.~\ref{ResonantPlots}.  \label{fig:paramrelic}}
\end{figure}

In Fig.~\ref{fig:paramrelic}, we show the allowed range of $(\mx,\mphi)$ with $\ax$ fixed by the relic density constraint as shown in Fig.~\ref{fig:relic}. For the vector mediator case (left), both attractive and repulsive interactions are present, and we take the average of attractive and repulsive cross sections.  In the scalar mediator case (right), DM self-interactions are purely attractive. It is clear that the allowed region for solving the small scale anomalies is still broad even after we impose the relic density constraint on $\ax$.

\section{Observational tests}

Self-interacting DM has distinct signatures in direct detection experiments because self-interactions thermalize the DM velocity distribution~\cite{Vogelsberger:2012sa}. In this section, we discuss signatures of self-interacting DM in indirect detection observations, when DM in halos self-annihilates. As we have shown, the existence of a light mediator is essential for generating a large enough self-scattering cross section. The same mediator can also lead to Sommerfeld enhancements for DM annihilation in halos if DM is symmetric. Since the enhancement effect increases as the DM velocity decreases, we expect DM particles in dwarf galaxies to have a larger self-annihilation cross section than those in the Milky Way or clusters. This scale-dependent feature of the DM annihilation cross section can be potentially determined by studying signal fluxes from different astrophysical objects.

Here, we take a few examples from the self-interacting DM models given in Section~\ref{sec:para} to show Sommerfeld enhancements for DM annihilation in halos. We consider the case where DM particles annihilate to SM states in DM halos with $s$-wave processes.\footnote{A familiar example is usual symmetric DM. Asymmetric DM can also generate annihilation signals if DM-anti-DM oscillations occur in the late epoch~\cite{Cohen:2009fz,Buckley:2011ye,Cirelli:2011ac,Tulin:2012re}.} To illustrate the point in a rather model-independent way, we take the assumption that DM has the correct relic density and do not demand $X\bar{X}\rightarrow\phi\phi$ to set the correct relic density as discussed in Section~\ref{sec:relic}. We have checked that our result does not change qualitatively if we demand the relic density set through $X\bar{X}\rightarrow\phi\phi$. 

For $s$-wave annihilation, the relative annihilation rates on different scales are determined by Sommerfeld enhancements folded together with DM distributions. Of course, DM self-interactions will also alter the density profiles in the center of the DM halos, changing the annihilation rates.   Rather than folding the DM distribution in to extract the total rate, we focus on the effect of the Sommerfeld enhancement alone on the annihilation cross section.  We calculate the thermally-averaged Sommerfeld enhancement factor as
\begin{eqnarray}
\left<S\right>= \int \frac{d^3 v}{(2\pi v_0^2)^{3/2}} \, e^{-\frac{1}{2} v^2/v_0^2}S_s,
\end{eqnarray}
where $S_s$ is the $s$-wave Sommerfeld enhancement factor given in Eq.~(\ref{eq:se}). 

\begin{figure}[t]
\includegraphics[scale=0.96]{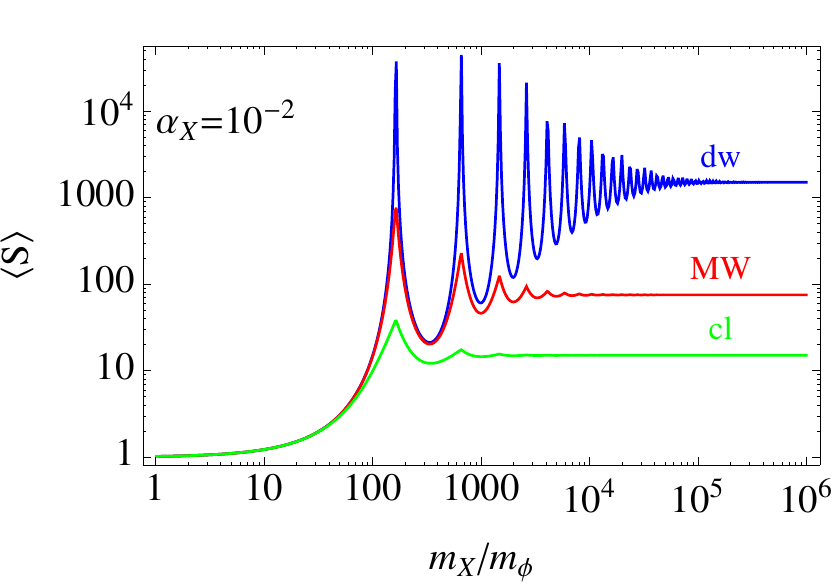}
\includegraphics[scale=0.96]{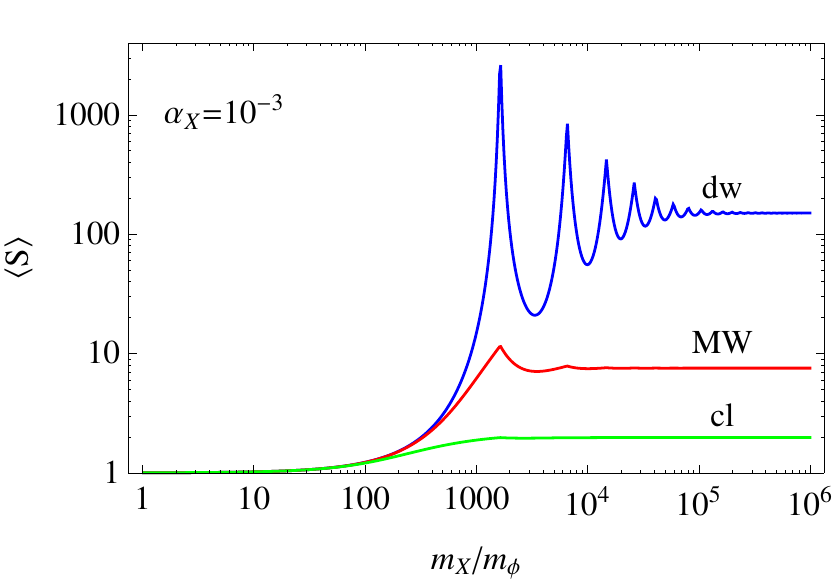}
\includegraphics[scale=0.64]{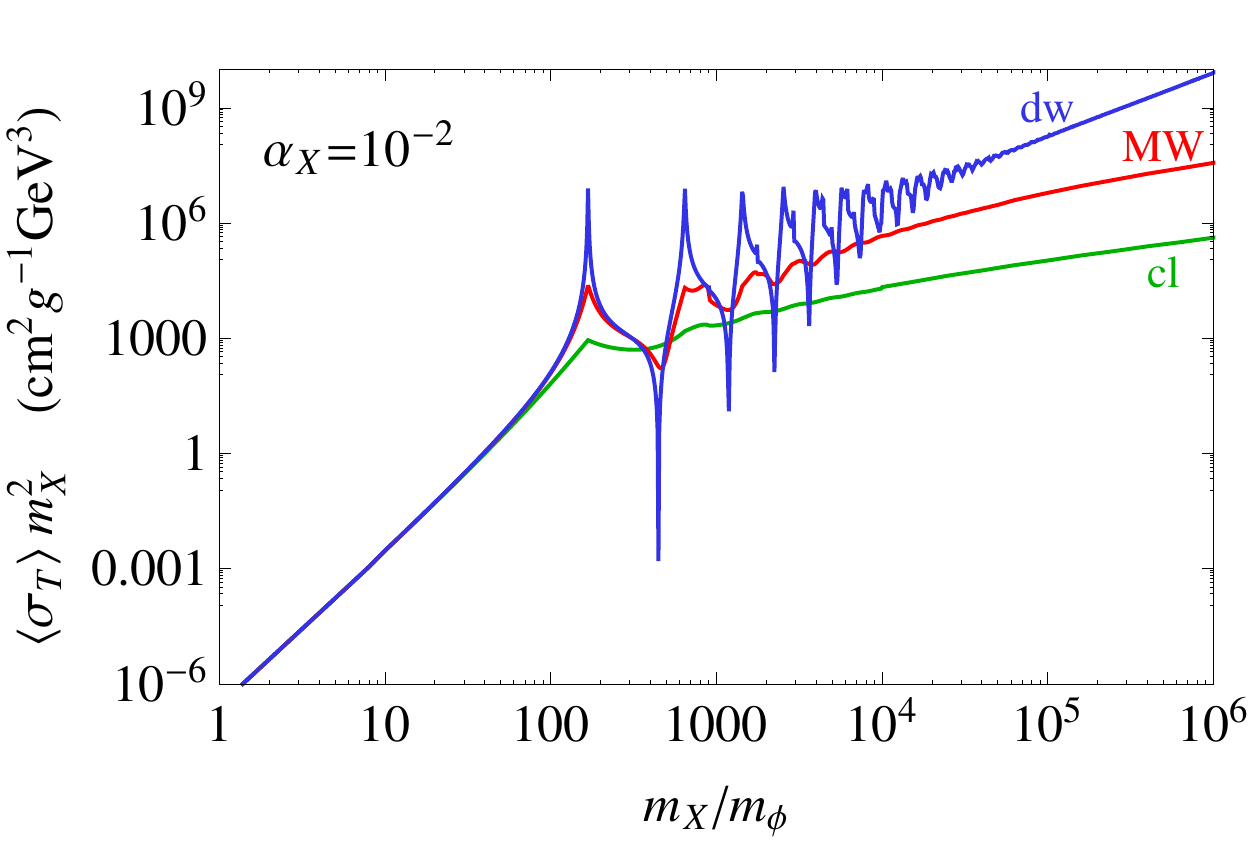}
\includegraphics[scale=0.64]{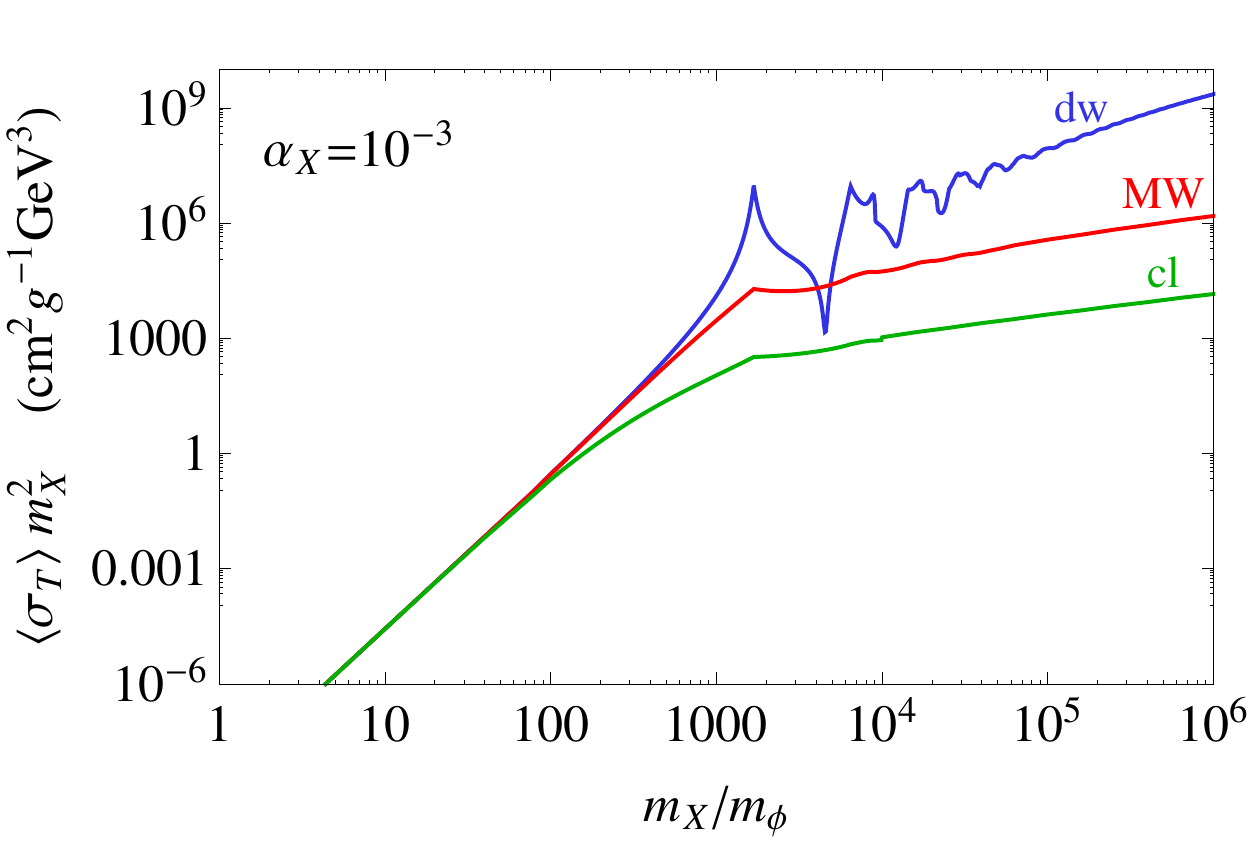}
\caption{The thermally-averaged $s$-wave Sommerfeld enhancement factor $\langle S \rangle$ and transfer cross section $\langle \sigma_T \rangle$ as a function of $m_X/m_\phi$ for $\ax=10^{-2}$ (left) and $\ax=10^{-3}$ (right) with $v_0=10~{\rm km/s}$ (blue), $200~{\rm km/s}$ (red) and $1000~{\rm km/s}$ (green), corresponding to the most probable DM velocities on dwarf (``dw"), Milky Way (``MW"), and cluster (``cl") scales.  One can see the correlation between the enhancement in the annihilation cross section and the scattering cross section due to the $s$-wave resonance. \label{fig:se}}
\end{figure}

In the top two panels of Fig.~\ref{fig:se}, we plot the thermally-averaged Sommerfeld enhancement factor for DM annihilation as a function of $\mx/\mphi$ for different $\ax$ and DM velocities. The upper two panels in Fig.~\ref{fig:se} are complementary to those in Fig.~\ref{ResonantPlots} and to the lower two panels of Fig.~\ref{fig:se}, which show the preferred parameter space for the self-scattering cross section to solve the small scale structure problem.   One can see the correlation between the enhancement in the annihilation cross section and the scattering cross section due to $s$-wave resonances.  It is also clear that, similar to the DM self-scattering case, there are three distinguishable regions for the Sommerfeld enhancement factor depending on $\mx/\mphi$. If the mediator and DM masses are comparable, it is in the Born regime where $\left<S \right>$ is negligible on all scales. On the other hand, if the DM mass is much larger than the mediator mass, the enhancement factor becomes independent of $\mx/\mphi$ which corresponds to the Coulomb  limit. In this limit, the enhancement factor is essentially given by $S\sim\pi\ax/v$. In the third region where $\mx/\mphi\simeq\pi^2n^2/6\ax$ with $n=1,2,3...$, DM annihilation can be enhanced resonantly.  On resonance, the enhancement factor is $S\sim\pi^2\ax\mphi/(6\mx v^2)$ which is very sensitive to the DM velocity. We emphasize that the $s$-wave resonant condition for the Sommerfeld enhancement of DM annihilation exactly corresponds to DM $s$-wave resonant self-scattering. 

As shown in Fig.~\ref{ResonantPlots}, most of the parameter space preferred for solving the small scale structure problem is in the resonant and classical regions. In these regions, constraints on the DM self-interacting cross section from DM halo shapes and the Bullet Cluster are elegantly evaded by the velocity-dependence of the self-scattering cross section. Interestingly, in the same regions, the Sommerfeld enhancements for the DM annihilation cross section differs significantly on different scales. In the resonant (classical) region, $\left<S\right>$ in dwarves can be a factor ${\cal O}(100)$ (${\cal O}(10)$) larger than that in the Milky Way. Therefore in many cases self-interacting DM predicts Sommerfeld enhancements for DM annihilation. If indirect detection signals are observed and annihilation cross sections are measured on different scales, it will give us a strong hint for self-interacting DM and help us further narrow down the parameter space.

\section{Conclusions}

We have examined DM self-interactions via a Yukawa potential with a massive dark force.  Over much of the parameter space, the Born ($\alpha_X m_X \lesssim m_\phi$) and classical ($m_X v \gtrsim m_\phi$) analytic formulae  break down, and quantum resonant structures, many with non-trivial velocity or angular dependences, arise.  We devised a method that allowed us to efficiently explore the strongly-coupled regime of parameter space.  We examined in detail the structure of this regime, and matched our results onto the known classical formula, verifying for the first time that analytical result.  We were also able to derive an analytic formula for our results for the case of a strongly-coupled $s$-wave resonance.  We also extracted the angular dependence of our results in the quantum and classical regimes, adding another dimension for study to the dynamics of DM self-scattering which is particularly important when the mediator is light. 

Our results have implications for the future study of DM self-interactions. Theoretical study and simulations of DM self-interactions have focused on simple analytic solutions for the scattering cross section, with constant or classical velocity (and no angular) dependence.    New simulations are in progress which will better account for baryonic effects on DM structure, while simultaneously integrating DM self-interactions \cite{privatecommunication}.  It will be important to simulate a broader class of DM self-interaction models by including strongly-coupled and resonant effects in the simulations.  Angular dependence should also be modeled, though including the general angular dependence in the strongly-coupled regime can be difficult.  However, we found a few cases where the scattering cross section has the desired velocity-dependence while the angular dependence is rather simple. In the case of $s$-wave resonant scattering, the scattering cross section scales as $v^{-2}$ and is also isotropic. In the strongly-coupled classical regime, we have numerically confirmed that isotropic assumption for scattering on dwarf galaxy scales which has been taken in the recent simulation~\cite{Vogelsberger:2012ku}. In addition, the Rutherford formula is available in the massless mediator limit.  We have devised benchmarks which may be utilized in simulations.  

In addition, our results allow the correlation of DM self-scattering with annihilation, having implications for indirect detection experiments. Sommerfeld enhancements for DM annihilation directly correspond to velocity dependent self-interacting DM.  Conversely, the absence of Sommerfeld enhancements imply a velocity-independent DM self-scattering cross section, so that if cores form in dwarves they also form in clusters. 

Clearly DM self-interactions provide an avenue for exploration with rich consequences for DM structure in our Universe.  While the nature of the DM may first be revealed through its interactions with ordinary matter, to date everything we have learned about DM has been gleaned through the formation of structure.  DM self-interactions can change this structure in complex ways, so that as we learn more about it,
we may also uncover evidence for the particle physics nature of DM.

\vspace{0.5cm}

{\em Acknowledgements:}  We thank F.~Governato, M. Kaplinghat, T.~Quinn, and S.~Tremaine for helpful discussions.  ST and KZ are supported by the DoE under contract de-sc0007859.  HBY and KZ are supported by NASA Astrophysics Theory Grant NNX11AI17G.  KZ is also supported by NSF CAREER award PHY 1049896.  

\appendix

%%%%%%%%%%%%%%%%%%%%%%%%%%%%%%%%%%
\section{Compendium of analytic results and benchmark points}
%%%%%%%%%%%%%%%%%%%%%%%%%%%%%%%%%5

We summarize analytic results for self-interacting DM scattering through a Yukawa potential.  The relevant parameters are the DM mass $m_X$, the dark force mediator mass $m_\phi$ and coupling $\alpha_X$, and the relative velocity $v$.  The transfer cross section $\sigma_T = \int d\Omega (1-\cos\theta) d\sigma/d \Omega$ provides a useful proxy for comparing specific particle physics models to N-body simulation results.  We also give $d \sigma/d \Omega$, which is a required particle physics input for simulations.

In the Born limit ($\alpha_X m_X/m_\phi \ll 1$), the cross section can be computed perturbatively in $\alpha_X$. The differential cross section is $d\sigma/d\Omega = \alpha_X^2 m_X^2/(m_\phi^2 + m_X^2 v^2 (1-\cos\theta)/2)^2$, giving
\begin{align} 
\sigma_T^{\rm Born} &= \frac{8\pi \alpha_X^2}{m_X^2 v^4} \Big( \log\big(1+m_X^2 v^2/m_\phi^2\big) -\frac{m_X^2 v^2}{m_\phi^2 + m_X^2 v^2} \Big) \label{born2} \; ,
\end{align}
for both attractive and repulsive potentials~\cite{Feng:2009hw}.

non-perturbative effects become important outside the Born regime ($\alpha_X m_X /m_\phi \gtrsim 1$).  Results have been obtained in the classical limit ($m_X v/m_\phi \gg 1$), giving for an attractive potential~\cite{Feng:2009hw,Khrapak:2003}
\beq
\sigma_T^{\rm clas} = 
\left\{\begin{array}{lc}
\frac{4 \pi}{m_\phi^2} \beta^2 \ln\left(1+\beta^{-1}\right) & \beta \lesssim 10^{-1} \\
\frac{8 \pi}{m_\phi^2} \beta^2 / \left(1+1.5 \beta^{1.65}\right) & \; 10^{-1} \lesssim \beta \lesssim 10^3 \\
\frac{\pi}{m_\phi^2} \left(\ln \beta+1-\frac{1}{2} \ln^{-1}\beta \right)^2 & \beta \gtrsim 10^3
\end{array} \right. \label{plasmaAppendix}
\eeq
and for a repulsive potential~\cite{Tulin:2012wi,Khrapak:2004}
\beq
\sigma_T^{\rm clas} = 
\left\{\begin{array}{ll}
\frac{2 \pi}{m_\phi^2} \beta^2 \ln\left(1+\beta^{-2}\right) & \beta \lesssim 1 \\
\frac{\pi}{m_\phi^2} \left(\ln 2 \beta-\ln \ln 2 \beta \right)^2 & \beta \gtrsim 1
\end{array} \right. \label{plasma2}
\eeq
where $\beta \equiv 2 \alpha_X m_\phi / (m_X v^2)$.  We find that $d \sigma/d\Omega \approx \sigma_T/(4\pi)$ (i.e., approximately constant) for $\beta \lesssim 1$, but approaches the Rutherford scattering formula $d \sigma/ d\Omega \approx \alpha_X^2/( m_X^2 v^4 \sin^4 \theta/2)$ for $\beta \gtrsim 1$.

Outside the classical regime ($m_X v/m_\phi \lesssim 1$), the cross section is largely dominated by $s$-wave scattering.  We have obtained a new exact non-perturbative result for $\sigma_T$ for the Hulth\'{e}n potential, which provides an excellent approximation for the true Yukawa potential.  Our result is:
\beq
\sigma_T^{\rm Hulth\acute{e}n} = \frac{16\pi}{m_X^2 v^2} \sin^2 \delta_0 \label{hulthen}
\eeq
where the $\ell=0$ phase shift is given in terms of the $\Gamma$-function by
\beq
\delta_0 = \arg\left( \frac{ i \,\Gamma\big( \frac{ i m_X v}{\kappa m_\phi} \big)}{\Gamma(\lambda_+) \Gamma(\lambda_-)} \right) \, , \quad
\lambda_\pm \equiv \left\{ \begin{array}{ll} 
1 + \frac{ i m_X v}{2 \kappa m_\phi} \pm  \sqrt{ \frac{ \alpha_X m_X}{\kappa m_\phi}  - \frac{ m_X^2 v^2 }{4 \kappa^2 m_\phi^2} } & {\rm attractive} \\
1 + \frac{ i m_X v}{2 \kappa m_\phi} \pm  i \sqrt{ \frac{ \alpha_X m_X}{\kappa m_\phi}  + \frac{ m_X^2 v^2 }{4 \kappa^2 m_\phi^2} } & {\rm repulsive} \end{array} \right.
\eeq
and $\kappa \approx 1.6$ is a dimensionless number.  The differential cross section is $d \sigma/d\Omega = \sigma_T/(4\pi)$.  This formula takes into account non-perturbative effects associated with $s$-wave scattering, and covers a complementary parameter region to the classical and Born formulae.

\end{document}